\documentclass[12pt,preprint]{aastex} 

\usepackage{multirow}
\usepackage{multicol}

\usepackage{natbib}
\usepackage{epsfig}
\usepackage{pdflscape}
\usepackage{subfigure}

\newcommand\arcpt{${{\lower3pt\hbox{$^{\prime\prime}$}}\atop{\raise4pt\hbox{.}}}$}

\slugcomment{submitted to {\it The Astronomical Journal} 20 April 2018}

\shorttitle{RECONS 10 Parsec Discoveries} 
\shortauthors{Henry et al.}

\begin{document}

\title{The Solar Neighborhood XLIV: RECONS Discoveries within 10 Parsecs}

\author{Todd J.~Henry$^{1,8}$,
        Wei-Chun Jao$^{2,8}$,
        Jennifer G.~Winters$^{3,8}$,
        Sergio B.~Dieterich$^{4,8}$,
        Charlie T.~Finch$^{5,8}$,
        Philip A.~Ianna$^{1,8}$,
        Adric R.~Riedel$^{6,8}$,
        Michele L.~Silverstein$^{2,8}$,
        John P.~Subasavage$^{7,8}$,   
        Eliot Halley~Vrijmoet$^{2}$
        \vskip15pt}


\vskip20pt

\affil {$^{1}$RECONS Institute, Chambersburg, PA 17201, USA; toddhenry28@gmail.com, philianna3@gmail.com}
\affil {$^{2}$Department of Physics and Astronomy, Georgia State University, Atlanta, GA 30302, USA; jao@astro.gsu.edu, silverstein@astro.gsu.edu, vrijmoet@astro.gsu.edu}
\affil {$^{3}$Harvard-Smithsonian Center for Astrophysics, Cambridge, MA 02138, USA; jennifer.winters@cfa.harvard.edu}
\affil {$^{4}$Department of Terrestrial Magnetism, Carnegie Institution for Science, Washington, DC 20015, USA; sdieterich@carnegiescience.edu}
\affil {$^{5}$Astrometry Department, U.S. Naval Observatory, Washington, DC 20392, USA; charlie.finch@navy.mil}
\affil {$^{6}$Space Telescope Science Institute, Baltimore, MD 21218, USA; adric.riedel@gmail.com}
\affil {$^{7}$United States Naval Observatory, Flagstaff, AZ 86001, USA; jsubasavage@nofs.navy.mil}   
\affil {$^{}$\vskip00pt}

\altaffiltext{8}{Visiting Astronomer, Cerro Tololo Inter-American
  Observatory.  CTIO is operated by AURA, Inc.\ under contract to the
  National Science Foundation.}


\vfil\eject

\begin{abstract}

We describe the 44 systems discovered to be within 10 parsecs of the
Sun by the RECONS team, primarily via the long-term astrometry program
at CTIO that began in 1999.  The systems --- including 41 with red
dwarf primaries, 2 white dwarfs, and 1 brown dwarf --- have been found
to have trigonometric parallaxes greater than 100 milliarcseconds
(mas), with errors of 0.4--2.4 mas in all but one case.  We provide
updated astrometric, photometric ($VRIJHK$ magnitudes), spectral type,
and multiplicity information here.  Among these are 14 systems that
are new entries to the 10 parsec sample based on parallaxes measured
at the CTIO/SMARTS 0.9m telescope.  These are the first parallaxes for
nine systems, while the remaining five systems had previously measured
parallaxes with errors greater than 10 mas or values placing them
beyond 10 parsecs.  We also present parallaxes from URAT for seven of
these systems, providing additional evidence that they are closer than
10 parsecs.  Among the 44 systems are 14 multiples, all of which are
promising targets for accurate mass determinations, including a few
with brown dwarf secondaries.  In addition, we provide new data for 22
systems that were previously known to lie within 10 parsecs and 9
systems reported to be closer than that horizon but for which new
parallaxes place them further away.  In total, we provide data for 75
systems, for which 71 have new or updated parallaxes here.

The 44 systems added by RECONS comprise one of every seven systems
known within 10 parsecs.  We illustrate the evolution of the 10 parsec
sample from the 191 systems known when the final Yale Parallax Catalog
(YPC) was published in 1995 to the 316 systems known today.  Even so
close to the Sun, additional discoveries of red and brown dwarfs (and
perhaps even white dwarfs) are likely, both as primaries and
secondaries, although we estimate that at least 90\% of the stellar
systems closer than 10 parsecs have now been identified.

\end{abstract}


\section{Introduction}
\label{sec:intro}

The solar neighborhood holds a special place in the human psyche
because, by our very nature, humans explore the nearest locales first.
Space is no exception.  The nearest stars provide the framework upon
which stellar astrophysics is based because the nearby star population
contains the most easily studied representatives of their kinds.  So,
it is essential that we make a careful reconnaissance of the nearest
stars to understand our Sun and Solar System in context.

The REsearch Consortium On Nearby Stars (RECONS, {\it www.recons.org})
was established in 1994 to discover ``missing'' members of the solar
neighborhood and to characterize the complete sample of nearby star
systems and their environs.  Over the past two decades, we have
published RECONS results in papers in this series, {\it The Solar
  Neighborhood}, in {\it The Astronomical Journal}.  This paper
focuses on the 10 parsec (pc) sample, in which a system comprised of
stars, brown dwarfs, and/or planets is included if a trigonometric
parallax, $\pi_{trig}$, of at least 100 mas has been determined, and
the parallax has a formal error, $\sigma_\pi$, of 10 mas or less.
Here we report trigonometric parallaxes measured at the CTIO/SMARTS
0.9m for 14 systems that are new additions to the 10 pc
sample,\footnote{We also provide parallaxes from URAT confirming that
  seven of the new systems are within 10 pc.} as well as for 30 other
systems discovered by RECONS to be within 10 pc.  We also outline the
growth of the 10 pc sample since the 1995 publication of the largest
ground-based compendium, {\it The General Catalogue of Trigonometric
  Stellar Parallaxes} \citep{vanAltena1995}, also known as the ``Yale
Parallax Catalog'' (hereafter YPC).  We include the contributions by
the space-based {\it Hipparcos} effort \citep{Perryman1997,ESA1997},
with initial results published in 1997, and utilize the updated
parallaxes from the \citet{vanLeeuwen2007} reduction done a decade
later; for stars within 10 pc, there are no significant differences
that change the sample membership.  The new results and overview
presented here set the stage for the full analysis of the 10 pc sample
to be evaluated after parallaxes are available from {\it Gaia} in Data
Release 2.

We outline the observing program and target sample in $\S$2, followed
by a description of the observations in $\S$3.  Astrometric,
photometric, and spectroscopic results for all 75 systems in this
study, including the 44 added by RECONS to the 10 pc sample, are given
in $\S$4.  Systems worthy of note are described in $\S$5.  Finally,
the evolution of the 10 pc sample since 1995 and the contributions
made by RECONS and others are placed in context in $\S$6.

\section{Observing Program and Sample}
\label{sec:sample}

A primary goal of the RECONS effort has been to discover and
characterize white, red, and brown dwarfs in the southern sky within
25 pc.  To this end, the RECONS team began a parallax program at Cerro
Tololo Inter-American Observatory (CTIO) in 1999 under the auspices of
the NOAO Surveys Program, using both the 0.9m and 1.5m CTIO/SMARTS
telescopes.  Optical photometry was also obtained at both telescopes,
and spectra were obtained at the 1.5m and CTIO 4.0m.  Parallaxes for a
total of 58 systems with $VRI$ = 11--23 were measured at the 1.5m and
published in \citet{Costa2005, Costa2006}.  The 0.9m effort has
continued to the present as an expanded astrometry/photometry program
via the SMARTS (Small and Moderate Aperture Research Telescope System)
Consortium, which was formed in 2003.  More than 800 systems with
$VRI$ = 9--18 have been observed for parallax at the 0.9m, with new
results presented here in $\S$4.

The objects targeted in the RECONS program at CTIO fall into two
general categories: those that have high proper motions or crude
trigonometric parallaxes indicating they might be nearby, and those
with optical/infrared photometry or spectroscopy hinting that they
might be closer than 25 pc.  Systems with declinations south of
$+$30$^\circ$ have been observed astrometrically, photometrically, and
spectroscopically to pinpoint their distances, proper motions,
absolute magnitudes, colors, and spectral types.  By providing this
full suite of data, well-characterized new members can be added to the
25 sample.  In this paper we collect and update results for the full
set of 44 new systems found within 10 pc via this effort, provide new
results for 22 previously identified 10 pc systems, and add results
for nine additional systems that were previously reported to be within
10 pc for which our parallaxes place them beyond that horizon.  In
total, we present information for 75 systems comprised of primarily
red dwarfs, with a few white and brown dwarfs as well.

\section{Observations}
\label{sec:observations}

\subsection{Astrometry}
\label{sec:observations.astr}

A single camera, equipped with a Tektronics 2048x2046 CCD having
pixels 401 mas square on the sky, has been used at the 0.9m since
1999.  The nearby systems described here were observed for 2.0--18.4
years (median coverage 13.1 years) in an effort to reach parallax
precisions of 3 mas or better, corresponding to distances accurate to
3\% at 10 pc.  Both astrometry and photometry observations (the latter
discussed in $\S$\ref{sec:observations.phot}) utilize only the central
quarter of the CCD (6.8\arcmin$\times$6.8\arcmin~field of view),
primarily to minimize astrometric distortions.  For the astrometric
observations, targets are observed through one of three $VRI$
filters\footnote{The central wavelengths for $V_{J}$, $R_{KC}$, and
  $I_{KC}$ are 5475\AA, 6425\AA~and 8075\AA, respectively, hereafter
  without the subscripts J $=$ Johnson and KC $=$ Kron-Cousins.},
except for a period from 2005 March through 2009 June, when the
cracked Tek \#2 $V$ filter was replaced by the very similar Tek \#1
$V$ filter.  Reductions indicate no significant differences in
photometric results from the two $V$ filters, although astrometric
offsets are seen, as described in \citet{Subasavage2009} and
\citet{Riedel2010}.  When possible, these astrometric offsets are
minimized by choosing reference stars very close to the targets.

A target's parallax and proper motion are usually considered worthy of
publication when $\sim$12 visits of $\sim$5 frames each are collected
on different nights spread over at least two years, typically
resulting in parallaxes with errors of $\sim$1.5 mas and proper
motions with errors $<$1 mas/yr.  The 75 systems described here have
parallax errors of 0.48--2.37 mas (median value 1.02 mas), except for
DEN 0255-4700 (error 3.98 mas), which was only observed at the 1.5m.
Thus, the resulting distances are known to $\sim$2\% or better.

As this is the 20th paper to report parallaxes from the
0.9m\footnote{The full library of papers can be found at {\it
    www.recons.org}.}, we provide only brief information on the
observing techniques and data reduction here; additional details can
be found in previous papers in this series, particularly in
\citet{Jao2005} and \citet{Henry2006}.  Data are reduced via IRAF with
typical bias subtraction and dome flat-fielding, using calibration
frames taken at the beginning of each night.  The frames are then
analyzed using our astrometric pipeline, utilizing images taken within
120 minutes of transit to produce parallaxes, proper motions, and
time-series photometry in the parallax filter (see
$\S$\ref{sec:results.phot}).  Target positions are measured relative
to 5--19 reference stars within a few arcminutes of the targets,
resulting in relative trigonometric parallaxes and proper motions.
Parallaxes are corrected from relative to absolute values using the
mean of the photometric distances to the reference stars.  For a few
targets, the distant reference stars appear reddened and result in
corrections of 3 mas or more.  In these cases we adopt our typical
correction of 1.5$\pm$0.5 mas.

We also report parallaxes from the United States Naval Observatory
Robotic Astrometric Telescope (URAT) in the notes for seven systems
($\S$\ref{sec:notes}) that support our RECONS measurements placing
these systems within 10 pc.  URAT observations were taken at CTIO over
a two-year period that overlapped with the long-term work described
for the RECONS effort, and details of the southern portion of the URAT
program can be found in \citet{Finch2018}.  Although the URAT parallax
errors of 3.4--5.6 mas for these seven systems are larger than for the
RECONS measurements, having two parallaxes that place systems within
10 pc provides valuable confirmation.

\subsection{Photometry}
\label{sec:observations.phot}

Optical $VRI$ photometry was acquired at the 0.9m using the same
camera/filter/detector combination used for the astrometry frames, as
described in $\S$\ref{sec:observations.astr}.  A modicum of additional
photometry was acquired at the SOAR 4.1m and CTIO/SMARTS 1.0m as part
of other RECONS programs (in particular, see \citet{Dieterich2014} and
\citet{Winters2011} for details).  All photometry is on the
Johnson-Kron-Cousins (JKC) system, determined using photometric
standards taken nightly to derive transformation equations and
extinction curves, primarily from \citet{Landolt1992}, supplemented
with standards from \citet{Bessell1990}, \citet{Graham1982}, and
\citet{Landolt2007,Landolt2013}.  These standard star sets were
augmented with a few of our own red standards --- GJ 1061, LHS 1723,
LHS 2090, SCR 1845-6357 AB, and SO 0253+1652 --- for which standard
$VRI$ magnitudes are included in Table 3.  The SOAR photometry was
taken using a Bessell filter set, but has been converted to the JKC
system as described in \citet{Dieterich2014}.


Photometric apertures 14\arcsec~in diameter were typically used to
determine the stellar fluxes to match the apertures used by Landolt.
Smaller apertures were used and aperture corrections done when targets
were corrupted by nearby sources such as physical companions or
background objects, or for very faint sources in an effort to reduce
the contribution of the sky background.  Typical errors in the $VRI$
values are 0.03 mag because longer exposure times are used for fainter
targets, as discussed in \citet{Henry2004} and \citet{Winters2011}.

\subsection{Spectroscopy}
\label{sec:observations.spec}

Spectra were obtained for nearby star candidates between 1995 and 2011
using the CTIO 1.5m and 4.0m telescopes.  We have previously reported
spectroscopy techniques and results in various papers in this series
\citep{Henry2002,Henry2004,Henry2006,Jao2008,Jao2011,Lurie2014,Riedel2011,Riedel2014}.
Briefly, at the 1.5m data were taken using a 2\farcs0~slit in the RC
Spectrograph with grating \#32 in first order and order blocking
filter OG570 to provide wavelength coverage from 6000--9500\AA~and
resolution of 8.6\AA~on the Loral 1200$\times$800 CCD.  At the 4.0m,
data were taken using a 2\farcs0~slit in the RC Spectrograph with
grating G181 and order blocking filter OG515 to provide wavelength
coverage from 5000--10700\AA~and resolution of 5.6\AA~on the Loral
3K$\times$1K CCD.  Typically, two sequential exposures were taken of a
star to permit cosmic ray removal.  Additional details of the
observational setups and data reduction can be found in
\cite{Henry2002} and \cite{Jao2008}.

Spectral types for many of the red dwarfs in the sample have been
reported in previous papers in this series.  We provide new spectral
types here, updated following the methodology described in
\cite{Riedel2014}.  Both the 1.5m and 4.0m data were reduced with
standard IRAF techniques using calibration frames of several flux
standards taken during each observing run.  Mild fringing in the 4.0m
data caused by back-illumination of the CCD was removed with a
tailored IDL routine.  After extraction, each spectrum was normalized
at 7500\AA, interpolated in 1\AA~intervals from 6000--9000\AA, and
best matches were found between the measured flux pairs for each
target and standards.  Telluric lines were not removed, so wavelengths
in telluric bands, as well as the region around the variable H$\alpha$
line, were omitted during matching.  The best spectral type was
selected using the lowest standard deviation value, measured using
$stddev\left( \frac{target}{standard} \right)$.  The resulting
spectral types for the M dwarfs in this study have errors of 0.5
subtypes.

The spectral types given here supersede those in previous papers,
e.g., types in \cite{Henry1994}, \cite{Henry2002}, and
\cite{Henry2006}, because the spectra have been reanalyzed on the new
system and consequent adjustments to types have been made.
Comparisons to spectral types from the large RHG survey of M stars
reported in \citet{Reid1995} and \citet{Hawley1996} indicate that the
types are consistent to $\sim$0.5 subtypes, with RECONS types being
slightly earlier in some cases.






\section{Results}
\label{sec:results}

\subsection{Astrometry}
\label{sec:results.astr}

Astrometry results are given in Table 1, with names in columns (1--2),
followed RA and DEC 2000.0 positions (3--4).  The next column lists
the discovery reference that added the system to the 10 pc sample via
$\pi_{trig}$ (5), and in cases of updates by us, we provide the
duration of the astrometric series at the time (6).  Specifics of the
observations are listed next, including the filter used (7), the
number of seasons (8), where ``c'' indicates continuous coverage and
``s'' indicates scattered coverage in at least some seasons, the
number of frames (9), the dates (10) and durations (11) of the series,
and the number of reference stars used in the reductions (12).
Measured relative trigonometric parallaxes ($\pi$ (rel), 13),
corrections to absolute parallaxes (14),\footnote{In four cases,
  generic corrections of 1.50 $\pm$ 0.50 mas have been used to adjust
  from relative to absolute parallaxes, as noted in the final column
  of Table 1.  This generic correction is used when the reference
  stars are reddened, resulting in erroneous distance estimates for
  stars comprising the reference field.} and absolute parallaxes
($\pi$ (abs), 15) are given next, followed by relative proper motion
magnitudes ($\mu$, 16) and position angles ($\theta$, north through
east, 17), and derived tangential velocities ($v_{tan}$, 18) using the
measured results.  In the final column (19) we provide brief notes.

The top portion of Table 1 lists the 44 systems that the RECONS effort
placed within 10 pc, with the proviso that the $\pi_{trig}$ error must
be less than 10 mas.  These 44 systems include 30 previously published
(most parallaxes updated here) and 14 new systems, indicated with
``new'' in column (5).  The 14 new systems added to the 10 pc sample
include nine singles and five doubles.  In the middle portion of Table
1 we list parallaxes for 22 additional important systems known to be
within 10 pc.  We have observed most of these for many years and in
some cases confirm parallaxes recently reported to be larger than 100
mas by other groups, whereas for other targets we provide updates that
differ by 5 mas or more from previously available values.  Finally,
the lower portion of Table 1 includes nine systems that were reported
to be within 10 pc, but our new parallaxes push them beyond that
horizon, including LP 647-013, LHS 1302, and APM 0237-5928 that we
previously had just within 10 pc.  These are to be contrasted with GJ
203 and GJ 595AB, two systems that were beyond 10 pc, but we pull into
the sample with new parallax measurements.  Noteworthy astrometric
results for individual systems are discussed in $\S$\ref{sec:notes}.


Of the 44 systems added to the 10 pc sample, six are within 5 pc,
including the closest, GJ 1061 (3.7 pc) that we first reported in
\cite{Henry1994}.  Six of the systems within 10 pc were discovered
during our SuperCOSMOS-RECONS (SCR) survey: SCR 0740-4257 (7.8 pc),
SCR 1546-5534 AB (9.7 pc), and SCR 2049-4012 AB (9.6 pc) have
parallaxes reported here for the first time, and we provide updated
measurements for SCR 0630-7643 AB (8.8 pc), SCR 1138-7721 (8.4 pc),
and SCR 1845-6357 AB (4.0 pc), all first reported in
\citet{Henry2006}.  The 44 systems have proper motions ranging from 68
to 5103 mas/yr, with 14 moving faster than 1\farcs0/yr.  Among these
are four of the five systems with the slowest proper motions within 10
pc: SCR 2049-4012 AB (68 mas/yr), WIS 0720-0846 AB (109 mas/yr), L
173-019 (126 mas/yr), and UPM 0815-2344 AB (138 mas/yr).


Figure \ref{fig:orbit} illustrates LHS 2090 as an example of a single
star with no detected companions, and five systems (shown
alphabetically by name) exhibiting perturbations with full orbits
covered in the astrometric data.  The photocentric positions for each
target have been separated into two panels, split into RA and DEC
components, and the points represent positions after shifts due to
proper motion and parallax have been removed.  Shown for the five
systems are orbital fits with periods of 0.17--6.74 years that have
been derived using the fitting protocol of \citet{Hartkopf1989}.  The
parallaxes given in Table 1 are those derived once the perturbations
have been removed.  In Table 2, we provide photocentric orbital
elements for these five systems, where for each we give the name,
number of orbits covered by the current dataset, orbital period ($P$),
in both years and days, photocentric semimajor axis ($a$),
eccentricity ($e$), inclination ($i$), longitude of periastron
($\omega$), argument of periastron ($\Omega$), and epoch of periastron
($T_0$).  Errors for each orbital element are given in the second row
for each system.  As is often the case with astrometric data, the
orbital periods are relatively well-constrained, particularly in the
cases of GJ 54 AB and GJ 595 AB, for which we have data spanning many
complete orbits.  The photocentric semimajor axes are known to
$\sim$10--25\% for the first four systems in Table 2.  However, for
each of the five orbits, the derived eccentricities, and consequently
the longitudes of periastron, are poorly determined using the current
data.  In the worst case, the orbital period for SCR 2049-4012 AB is
likely to be reliable, but because the derived inclination is nearly
edge-on, we deem the rest of the elements to be suspect; additional
data will be required to derive a useful orbit.  Comments on these
five systems with orbits are given in $\S$\ref{sec:notes}.


Figure \ref{fig:perts} illustrates six more systems showing
astrometric perturbations for which we do not yet have data covering
complete orbits.  For these systems, it is clear that close companions
are present, but the orbital elements remain uncertain, so we do not
provide them here.  By far the largest perturbation is that of WT 460
AB, which shows shifts by more than 250 mas in both RA and DEC caused
by a low mass stellar companion at a separation of $\sim$500 mas.
Comments on each of these six systems are also given in
$\S$\ref{sec:notes}.

%

%

\subsection{Photometry}
\label{sec:results.phot}

Photometry results in the $VRIJHK$ filters are given in Table 3, where
variability measurements are also listed.  Some of our $VRI$
photometry has been published in previous papers in this series (e.g.,
Winters et al.~2015), but here we list RECONS values with the number
of nights of data rather than the individual references, including
updated values when additional nights of $VRI$ data have been
acquired.  Thus, values given here supersede previously published
values for these objects.

Table 3 lists names for the systems (column 1), coordinates (2,3),
$VRI$ photometry (4,5,6) and the number of observations (7), where the
three digits indicate the numbers of nights in which magnitudes were
measured in the $VRI$ bands, respectively.  Note that fainter objects
sometimes have limited $V$ measurements.  The next columns give the
filter in which the images used for both astrometry and variability
were taken (8) and the variability in millimagnitudes in that filter
(9).  These results are followed by the 2MASS $JHK$ magnitudes
(10,11,12), the spectral types, luminosity classes, and references
(13,14,15), and notes (16).  The indicator ``J'' beside $VRIJHK$
photometric values and spectral types is used to represent joint
measurements of two or more components.


The level of variability in millimagnitudes (mmag) in the parallax
filter listed in column (9) of Table 2 has been derived as described
in previous papers in this series, in particular, \citet{Jao2011},
\citet{Hosey2015}, and \citet{Clements2017}.  The method outlined in
\citet{Honeycutt1992} has been used to derive the nightly offsets and
zero points for relative instrumental photometry to derive stellar
variability.

Among the 75 components with variability measurements, 11 vary by more
than 20 mmag (2\%), our adopted cutoff for a clearly variable star, as
described in \citet{Hosey2015}.  We interpret variability above 20
mmag to be due to spots that move in and out of view as a star rotates
on short timescales and/or that come and go over decades in the form
of stellar spot cycles.  Note that even below 20 mmag, some stars show
evidence of clear brightness changes, at least down to 15 mmag in our
data.  Light curves for 12 of the most compelling cases are shown in
Figure \ref{fig:cycle} and Figure \ref{fig:trend}, and can be broken
into three subsets:

{\bf Cycles:} Remarkable among the observed stars are several that
show clear evidence of long-term stellar cycles.  Light curves for six
stars are shown in Figure \ref{fig:cycle}, with cycles that last 7--15
yr.  The clearest example is GJ 1128 (20.5 mmag in $V$) with a 5-yr
cycle for which we see more than two cycles.

{\bf Trends:} The top four panels of Figure \ref{fig:trend} show light
curves for stars exhibiting long-term trends that may be portions of
cycles extending beyond the current datasets.  In particular, the
closest pre-main sequence star \citep{Riedel2011} AP Col (22.9 mmag in
$V$) shows evidence of a brightening trend of nearly 10\% from
2007-2018 due to a long-term cycle, upon which are superimposed
significant short-term variations presumably due to spots.  WT 460 AB
(18.8 mag in $I$) also shows a clear trend, having grown fainter by
6\% from 2000-2013 and brightening since.

{\bf Spots:} The bottom two panels of Figure \ref{fig:trend} show
light curves for the two most extreme cases of variability in the
sample --- the close binary GJ 867 CD (31.5 mmag at $V$), which has an
orbital period of 1.8 days (see $\S$\ref{sec:notes}), and Proxima
Centauri (36.5 mmag at $V$).  In fact, with the largest variability
measurement in the sample, Proxima Centauri shows substantial changes
in brightness throughout the 17 years of observations.


%
%
%
%
%
%
%
%

\subsection{Spectroscopy}
\label{sec:results.spec}

Spectral types and luminosity classes are given in Table 3, along with
references.  Nearly all of the systems in this paper have primaries
that are M dwarfs for which we provide updated types, indicated with
asterisks in column (15) of Table 3.  As described in
$\S$$\ref{sec:observations.spec}$, the M types presented here
supersede types given in previous papers in this series.  For the few
stars that we did not observe, types are taken from the literature, or
for those that have no spectroscopic observations, we estimate the
type based on the star's $V-K$ color and give ``estim'' as the
reference.  As with the photometry, a ``J'' is listed (in the
luminosity class column) when the light of two or more close
components was merged in the spectrum used for typing.

Among the 44 systems added by RECONS there are 31 with types M3V--M5V,
as is typical for the solar neighborhood population.  There are a few
objects that are not main-sequence M dwarfs.  The pre-main sequence
star, AP Col, is noted as luminosity class ``pms'' rather than ``V''.
Types for the two white dwarfs are DQpec for WD 0038-226 from
Giammichele et al. (2012), and DAZ for WD 0141-675 from Debes \& Kilic
(2010); further details about the white dwarf spectral types can be
found in \citet{Subasavage2017}.




\section{Systems Worthy of Note}
\label{sec:notes}

Here we provide additional details for many of the systems presented
in this paper, including the 14 new 10 pc systems and each of the 17
multiple systems.  For ease of use, all notes on systems are collected
in this section and are listed by RA (in italics).  In the
descriptions below, our photometric distance estimates are from two
sources.  We use SuperCOSMOS $BRI$ magnitudes derived from
photographic plate scans combined with 2MASS $JHK$ to estimate
distances accurate to 26\%, as described in \citet{Hambly2004}.
Similarly, we use our own $VRI$ photometry, primarily from the 0.9m,
combined with 2MASS $JHK$ to estimate distances accurate to 15\%, as
described in \citet{Henry2004}.  In particular, offsets between these
photometrically estimated distances and the trigonometric distances
often reveal flux contributed by unseen companions.


$\bullet$ {\it 0024$-$2708} : {\bf GJ 2005 ABC} first entered the 10
pc sample in \citet{Costa2005} with $\pi_{trig}$ = 129.47 $\pm$ 2.48
mas based on 2.3 years of 1.5m data, targeted primarily because the
YPC value of $\pi_{trig}$ = 135.3 $\pm$ 12.1 mas had an error larger
than 10 mas.  From the 0.9m we present a new value of $\pi_{trig}$ =
139.30 $\pm$ 1.59 mas using 18.3 years of data after taking into
account the notable curvature due to orbital motion of BC around A
shown in Figure \ref{fig:perts}.  As a check of the new larger
parallax, a reduction using only data from 2013 to 2016 was done to
minimize the effect of the orbital curvature, and a consistent
parallax was found.  The orbital fit is only a guide given that the
orbital period of BC around A is likely a century or more, while the
BC pair has an orbital period of 16.1 yr \citep{Leinert2001}.  With
masses of 0.07--0.08 $M_{\odot}$ (see Benedict et al.~2016), the B and
C components are crucial to our understanding of the transition region
between stars and brown dwarfs.  The combined light from the three
components exhibits a variation in the $R$ band with period $\sim$13
years.


$\bullet$ {\it 0109$-$0343} : {\bf LP 647-013} was reported in
\citet{Costa2005} to have $\pi_{trig}$ = 104.23 $\pm$ 2.29 mas from
our program at the CTIO 1.5m, using data spanning 3.2 years.
\citet{Weinberger2016} recently measured $\pi_{trig}$ = 97.55 $\pm$
2.04 mas, and when combined with our new value from the 0.9m program,
$\pi_{trig}$ = 94.11 $\pm$ 0.85 mas using data spanning 11.0 years,
this M9.0V star is pushed beyond 10 pc.

\begin{figure}[ht!]
  \subfigure{\includegraphics[scale=0.28,angle=00]{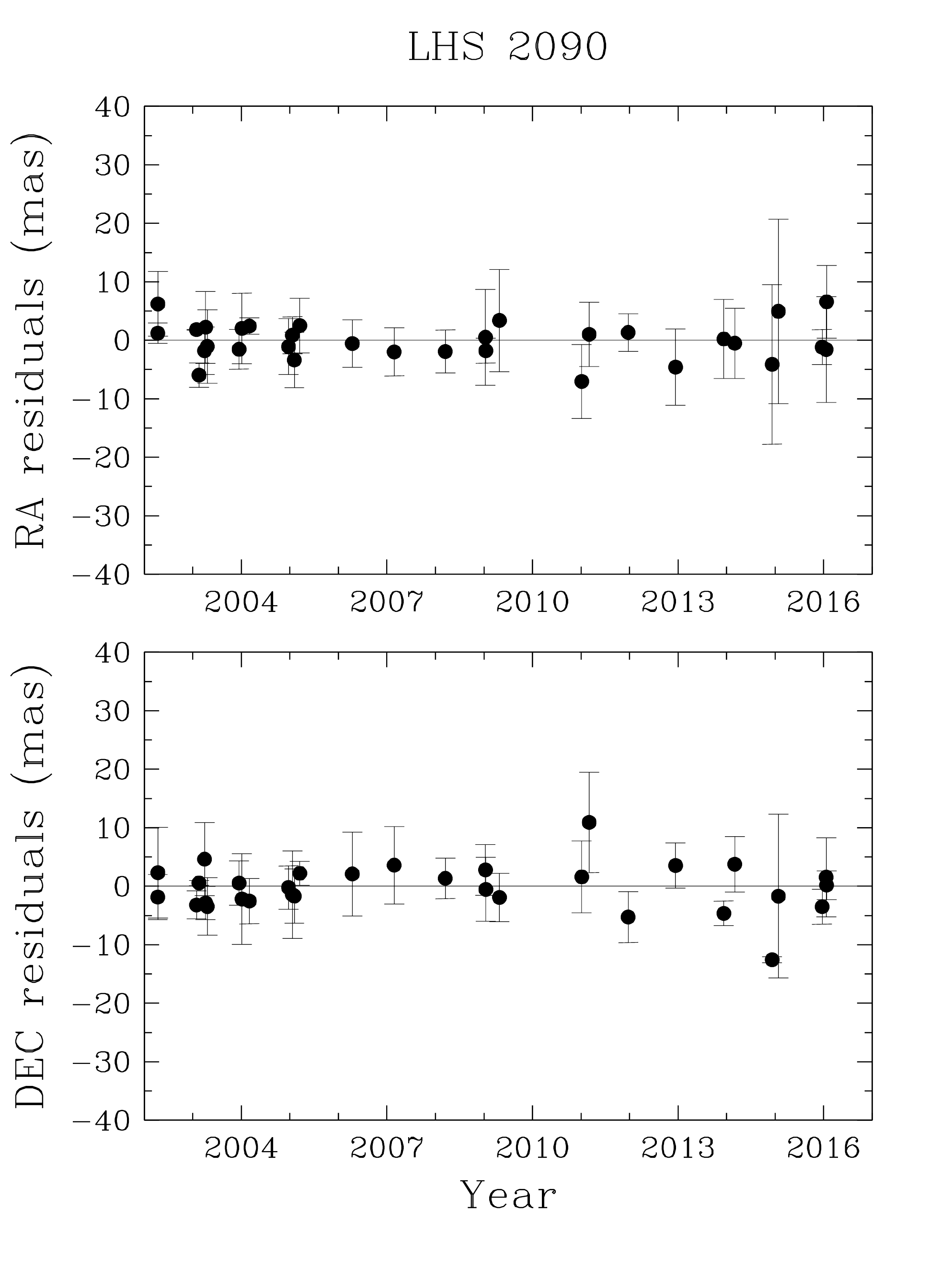}}
\hspace{-0.16in} 
 \subfigure{\includegraphics[scale=0.28,angle=00]{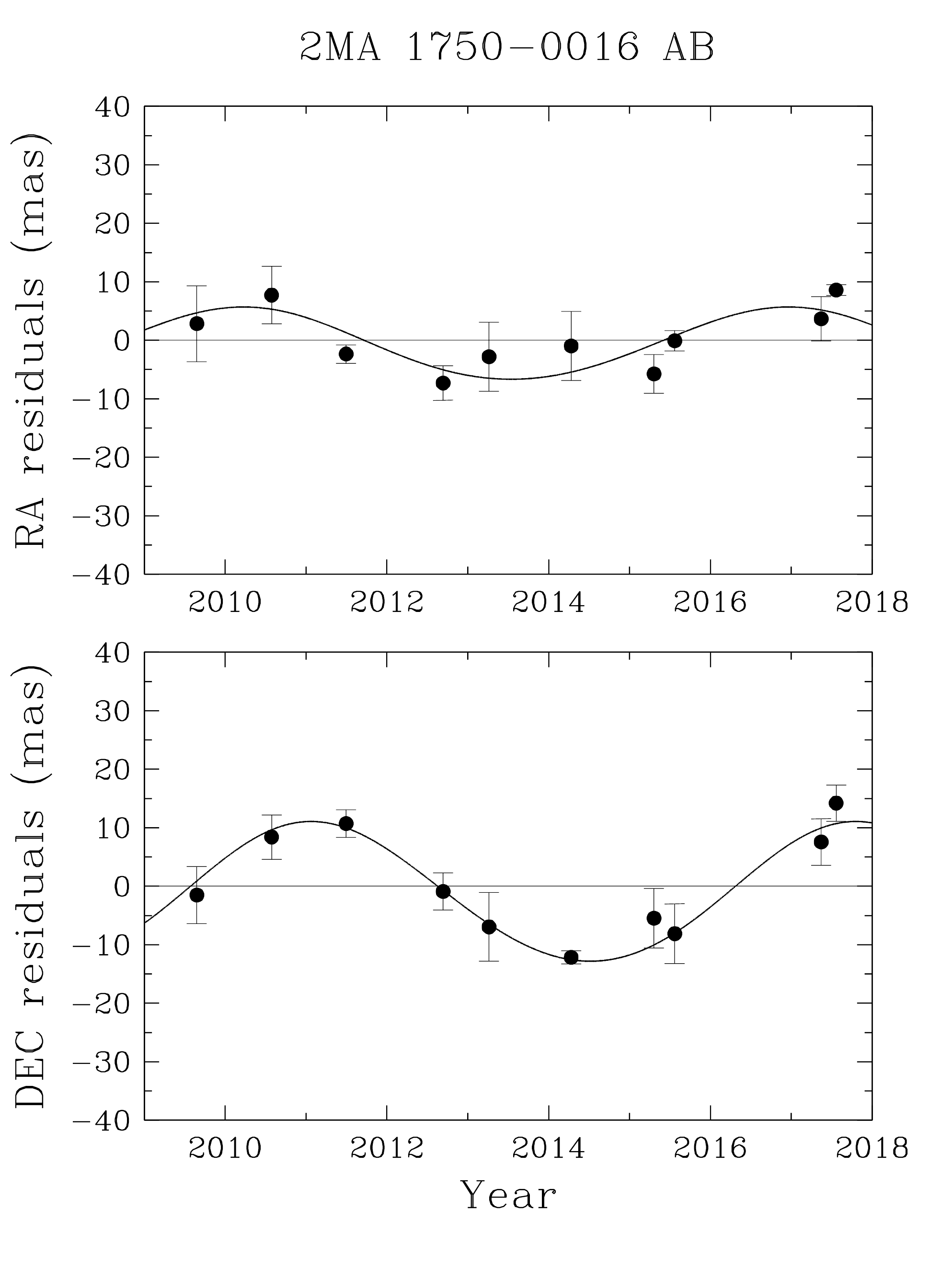}}
\hspace{-0.16in} 
  \subfigure{\includegraphics[scale=0.64,angle=00]{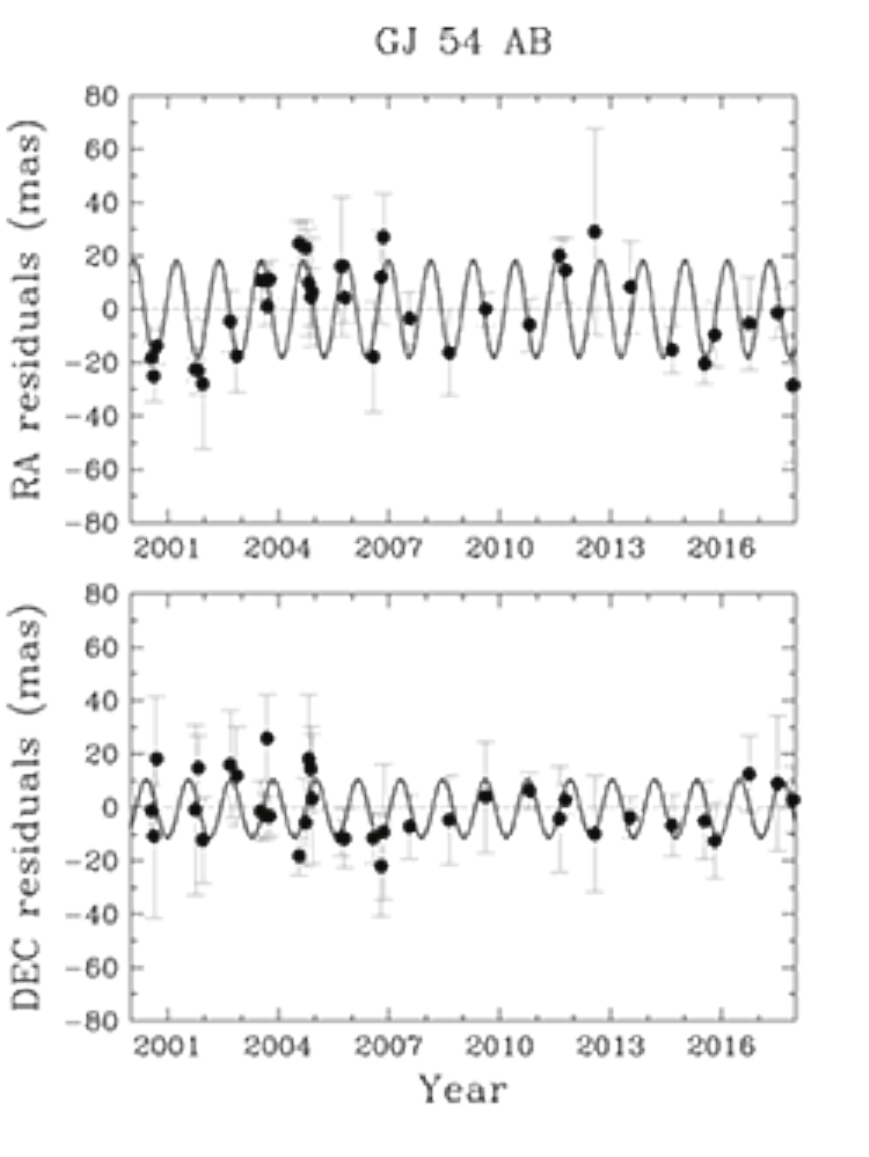}}
\hspace{-0.16in} 

\vskip-00pt
\subfigure{\includegraphics[scale=2.5,angle=00]{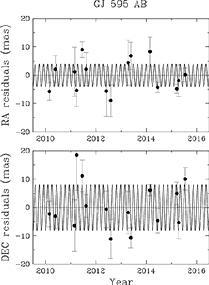}}
\hspace{-0.06in}
\subfigure{\includegraphics[scale=0.28,angle=00]{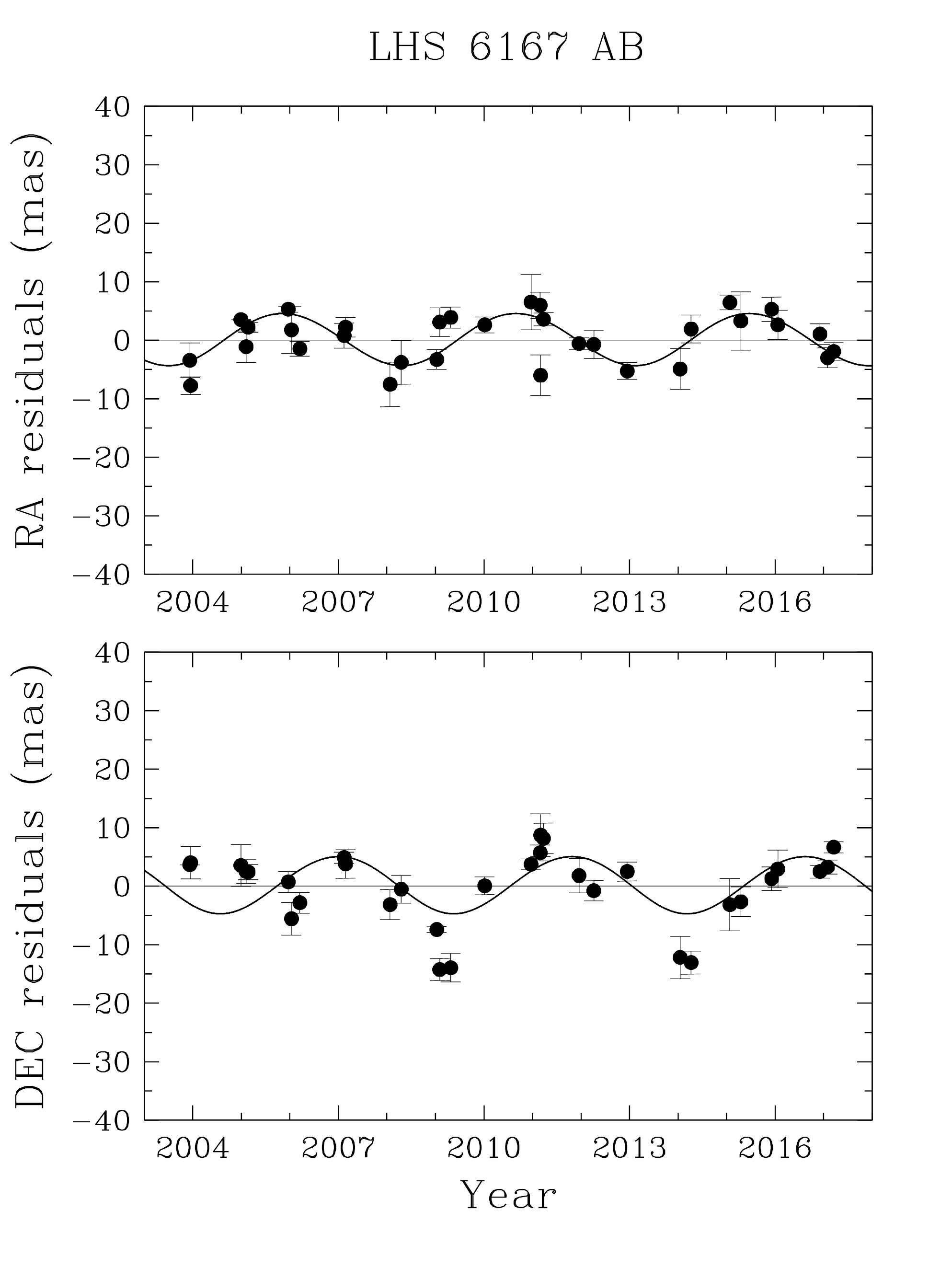}}
\hspace{-0.16in}
\subfigure{\includegraphics[scale=2.6,angle=00]{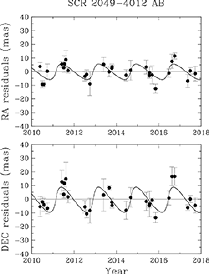}}

\vskip-10pt

\caption{\scriptsize Astrometric residuals in milliarcseconds in the
  RA and DEC directions are shown for six systems, after solving for
  parallax and proper motion.  Each point represents typically five
  frames taken on a single night.  Note the different vertical scales
  that span $\pm$20 mas to $\pm$80 mas.  The upper left panel shows
  flat residuals for the single star LHS 2090.  The remaining panels
  illustrate fits to the residuals for five multiples with orbits that
  have wrapped in our datasets; orbital parameters are given in Table
  2. \label{fig:orbit}}

\end{figure}

\begin{figure}[ht!]

{\includegraphics[scale=0.28,angle=00]{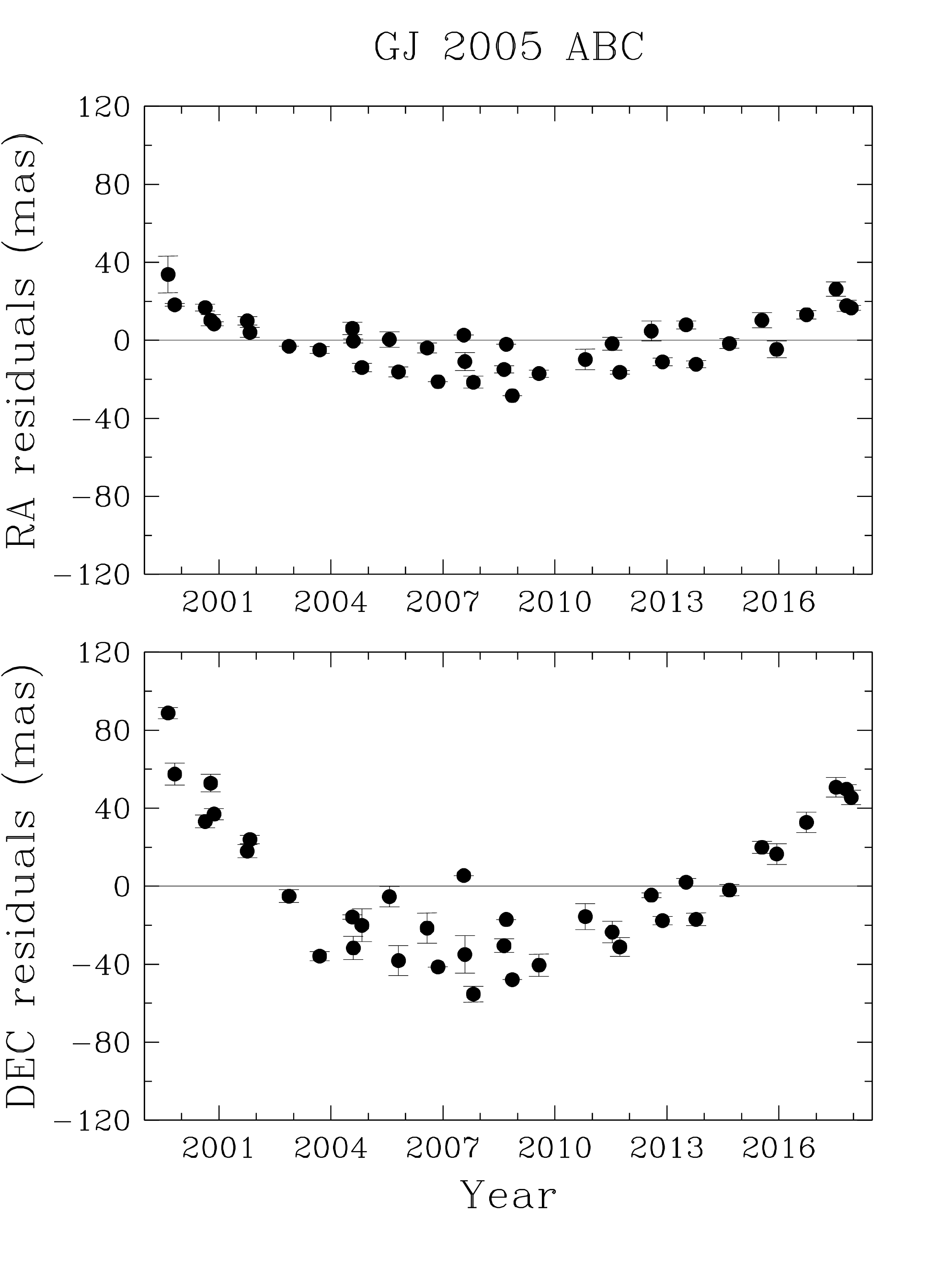}}
\hspace{-0.16in}
\subfigure
{\includegraphics[scale=0.28,angle=00]{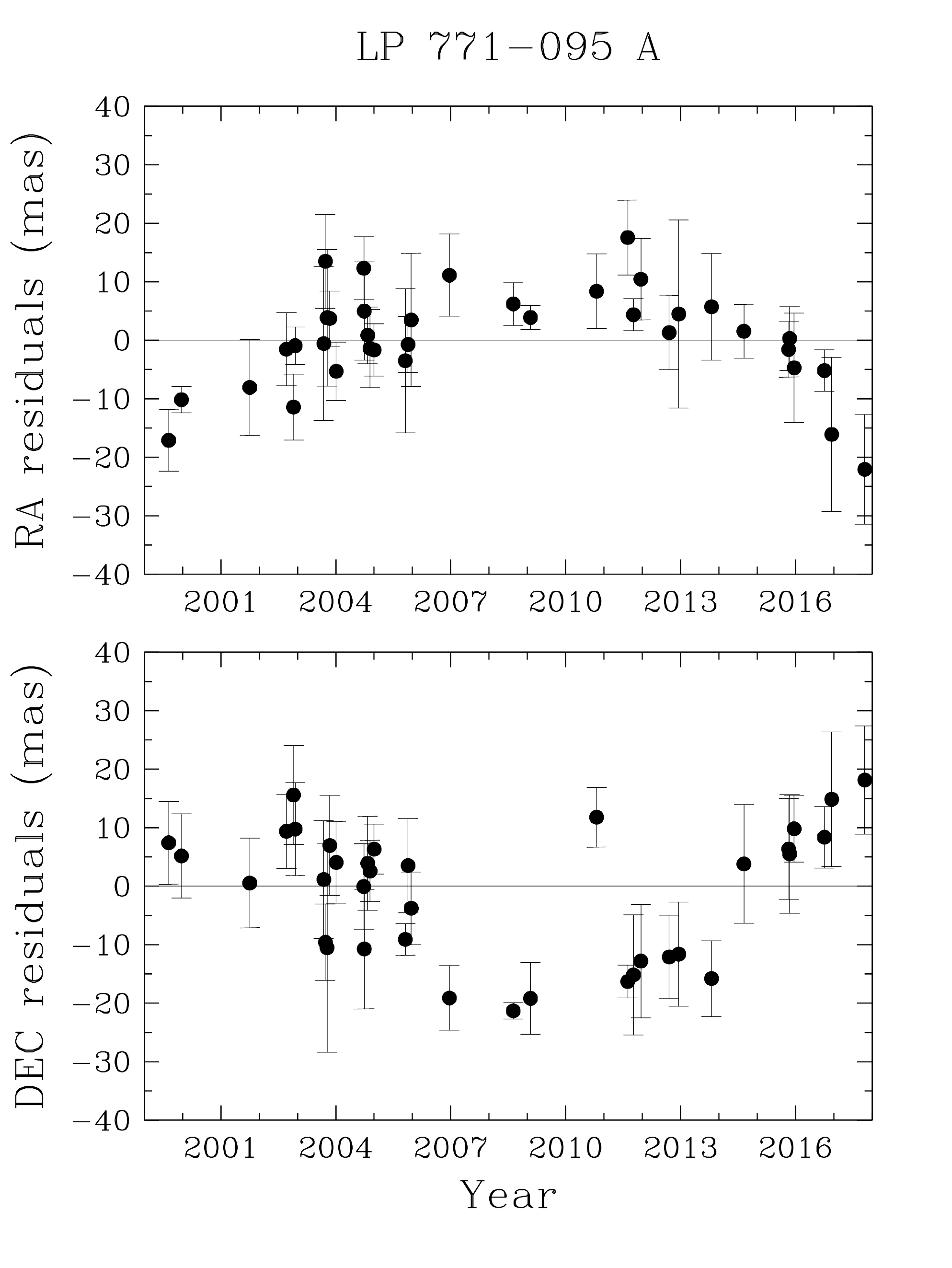}}
\hspace{-0.16in}
\subfigure
{\includegraphics[scale=0.28,angle=00]{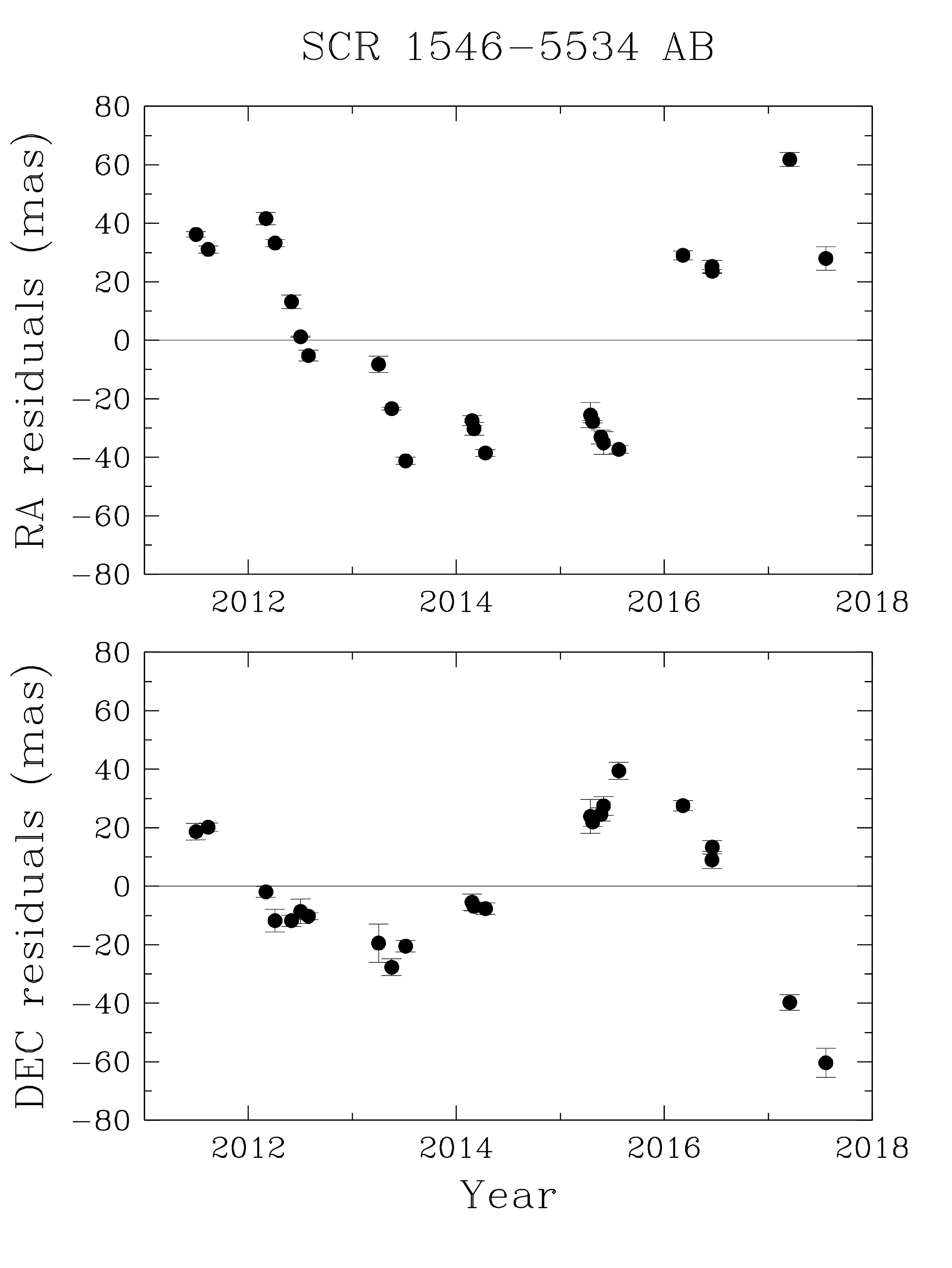}}

\vskip-00pt

{\includegraphics[scale=0.28,angle=00]{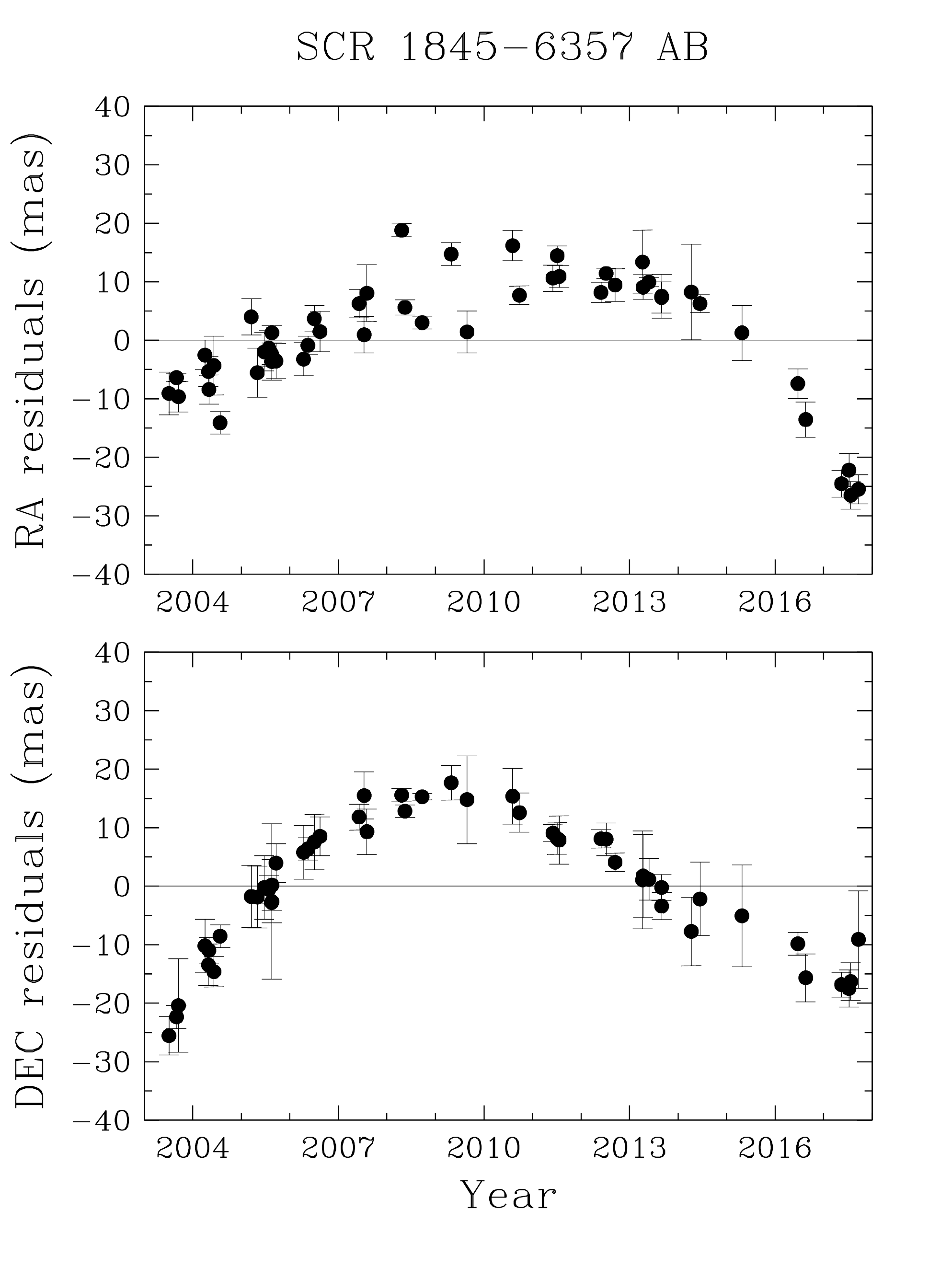}}
\hspace{-0.16in}
\subfigure
{\includegraphics[scale=0.28,angle=00]{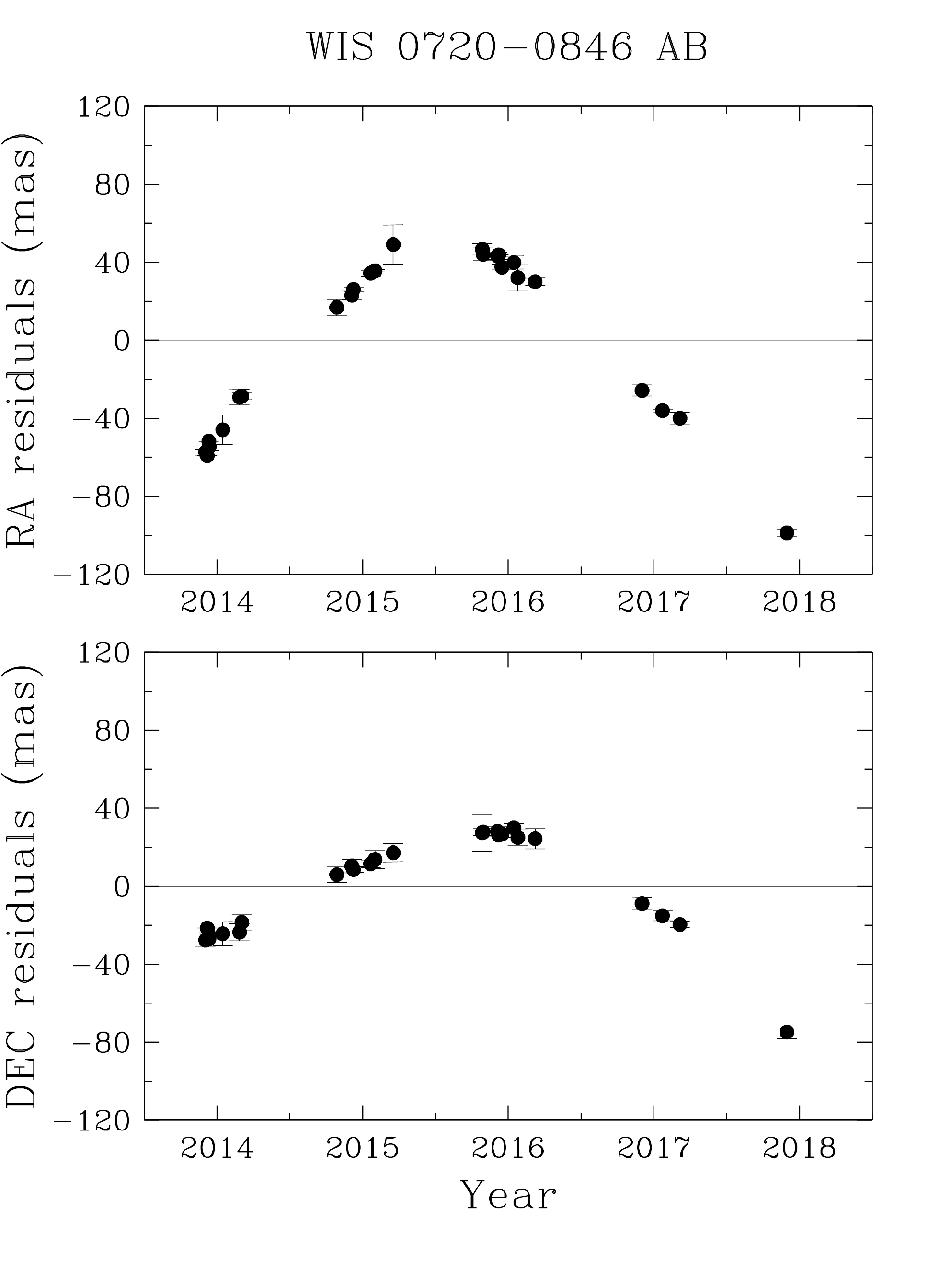}}
\hspace{-0.16in}
\subfigure
{\includegraphics[scale=0.28,angle=00]{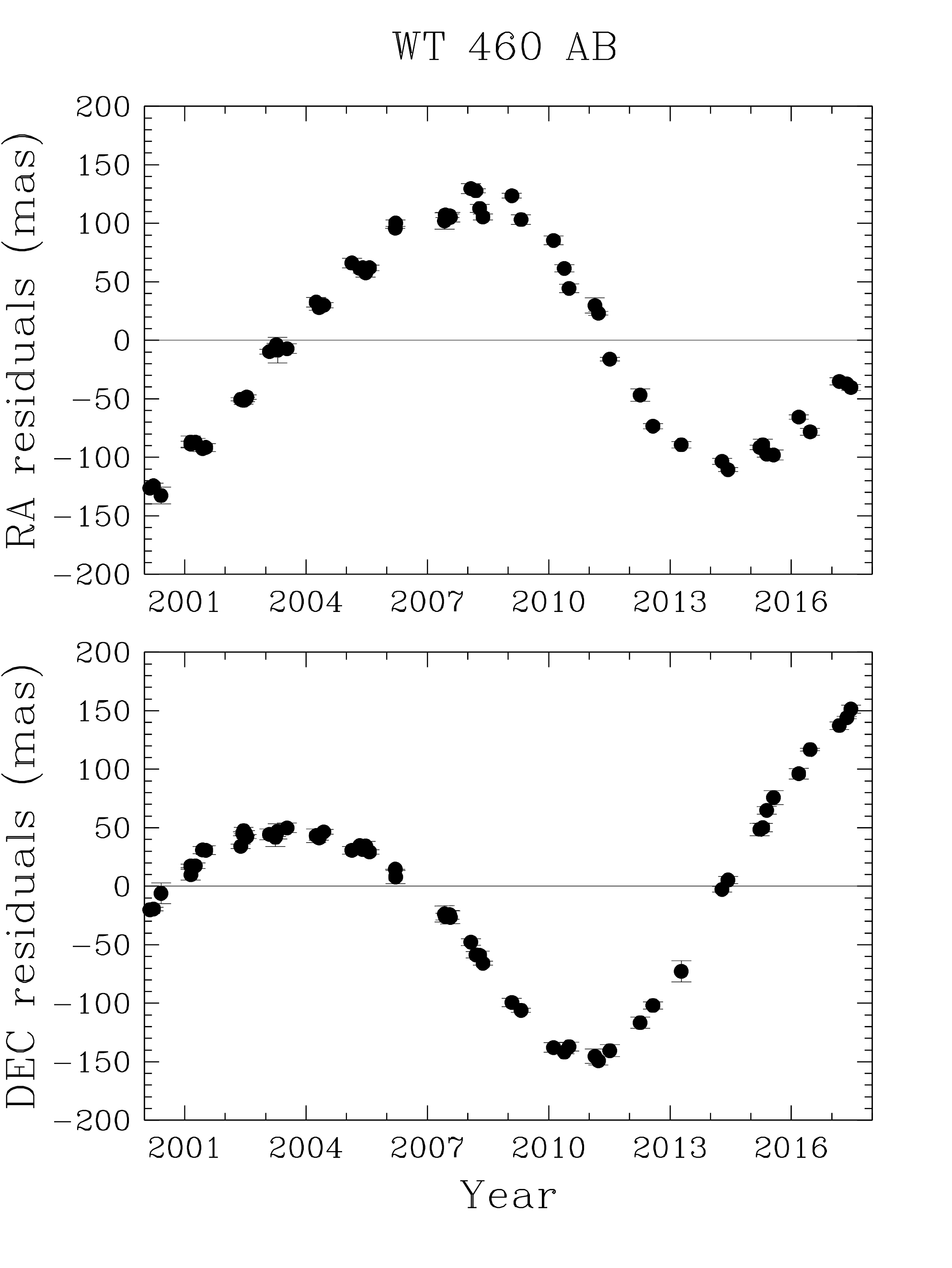}}

\vskip-10pt

\caption{\scriptsize Astrometric residuals in milliarcseconds in the
  RA and DEC directions are shown for six systems, after solving for
  parallax and proper motion.  Each point represents typically five
  frames taken on a single night.  Note the different vertical scales
  that span $\pm$40 mas to $\pm$200 mas.  All six systems show clear
  evidence of photocentric perturbations, but none have yet wrapped
  into closed orbits in our datasets. \label{fig:perts}}

\end{figure}

$\bullet$ {\it 0110$-$6726} : {\bf GJ 54 AB} was suspected to be a
binary by \citet{Rodgers1974} and was confirmed to be a close pair of
red dwarfs by \citet{Golimowski2004}, who imaged the secondary using
$HST$'s Fine Guidance Sensors (FGS) and $HST$'s Near Infrared Camera
and Multi-Object Spectrometer (NICMOS).  Continuing work by RECONS
using $HST$-FGS indicates a parallax of 126.9 $\pm$ 0.4 mas, an
orbital period of 418.5 days (1.15 yr), semimajor axis of the relative
(not photocentric) orbit of 126.4 mas, $\Delta$$V =$ 1.04 mag, and
masses of 0.43 and 0.30 $M_{\odot}$ \citep{Benedict2016}.  The system
provides a classic case of orbit entanglement that wreaks havoc on
parallax measurements because the orbital period nearly matches the
1.00 yr period of parallactic motion.  Although observed using faint
reference stars and only averaging 20 seconds per integration, after
more than 15 orbital cycles covered in our astrometry program over
17.4 years, we fit for the orbital motion (see Table 2 and Figure
\ref{fig:orbit}), finding $P_{orb}$ = 419.1 $\pm$ 1.3 days, consistent
with the FGS orbit.  The resulting $\pi_{trig}$ = 124.94 $\pm$ 2.08
mas is more reliable than the early value in \cite{Henry2006}, which
was undoubtedly corrupted by the orbital motion.  The matching
parallaxes from the 0.9m and $HST$ indicate a semimajor axis for the
orbiting pair of almost precisely 1.00 AU.

$\bullet$ {\it 0151$-$0607} : {\bf LHS 1302} was reported in
\citet{Henry2006} to have $\pi_{trig}$ = 100.78 $\pm$ 1.89 mas,
placing it near the 10 pc border.  Additional data extend the coverage
from 6.2 years to 15.2 years, resulting in a new value of $\pi_{trig}$
= 97.03 $\pm$ 0.92 mas (10.3 pc).  The star was reported in
\citet{Riedel2014} to be a potential kinematic match to the $\sim$30
Myr old Columba association, although its surface gravity and
isochrone placement suggest it is a field object.  It is not
noticeably elevated from the main sequence, but was observed to flare
by 0.15 mag in the $R$ band on UT 2004 November 22.


$\bullet$ {\it 0200$-$5558} : {\bf L 173-019} is a new low proper
motion ($\mu$ = 126 mas/yr) member of the solar neighborhood at only
8.1 pc.  This is one of four new systems reported here with proper
motions less than 200 mas/yr, ranking fourth slowest within 10 pc.
The reference stars are faint compared to L 173-019 ($V$ = 11.90),
resulting in a relatively large parallax error (2.08 mas).  Our value
of $\pi_{trig}$ = 123.11 $\pm$ 2.08 mas places the star comfortably
within 10 pc, and URAT's $\pi_{trig}$ = 112.1 $\pm$ 3.4 mas supports
the new 10 pc membership.

$\bullet$ {\it 0237$-$5928} : {\bf APM 0237-5928} was reported in
\citet{Henry2006} to have $\pi_{trig}$ = 103.72 $\pm$ 1.12 mas,
placing it near the 10 pc horizon.  With a dataset that now stretches
nearly three times longer (18.1 years), we find $\pi_{trig}$ = 99.75
$\pm$ 0.85 mas, placing the system just beyond 10 pc.

$\bullet$ {\it 0253$+$1652} : {\bf SO 0253+1652} was discovered by
\citet{Teegarden2003} as a potential nearby star of very high proper
motion with $\mu$ = 5.1$\arcsec$/yr and $\pi_{trig}$ = 410 $\pm$ 90
mas (2.4 pc).  We reported the first accurate parallax of 260.63 $\pm$
2.69 mas in \citet{Henry2006} using 2.4 years of astrometry data,
which was confirmed by \citet{Gatewood2009}, who found $\pi_{trig}$ =
259.25 $\pm$ 0.94 mas.  Here we update our parallax to $\pi_{trig}$ =
261.01 $\pm$ 1.76 mas (3.8 pc) using 12.3 years of data.  The system
currently ranks as the 24th closest to the Sun.

$\bullet$ {\it 0301$-$1635} : {\bf LTT 01445 ABC (LP 771-095 and LP
  771-096 AB)} is a nearby triple first reported to be within 10 pc in
\citet{Henry2006}, with $\pi_{trig}$ = 146.39 $\pm$ 2.92 mas for A and
139.70 $\pm$ 4.99 mas for BC based on 6.3 yr of data.  The BC pair is
separated by 1\farcs3 and is 7$\arcsec$ from A.  Astrometric residuals
for A show the pull by the BC pair shown in Figure \ref{fig:perts},
and this trend has been removed in the reported parallax.  We detect
orbital motion in the BC pair, and an orbital fit with long period has
been used to reduce the trend in the residuals, but the fit is not yet
reliable.  After 18.2 yr of data, the parallaxes are consistent and
the errors have been cut roughly in half, to 143.85 $\pm$ 1.59 mas for
A and 142.57 $\pm$ 2.03 mas for BC.

\begin{figure}[ht!]

{\includegraphics[scale=0.32,angle=270]{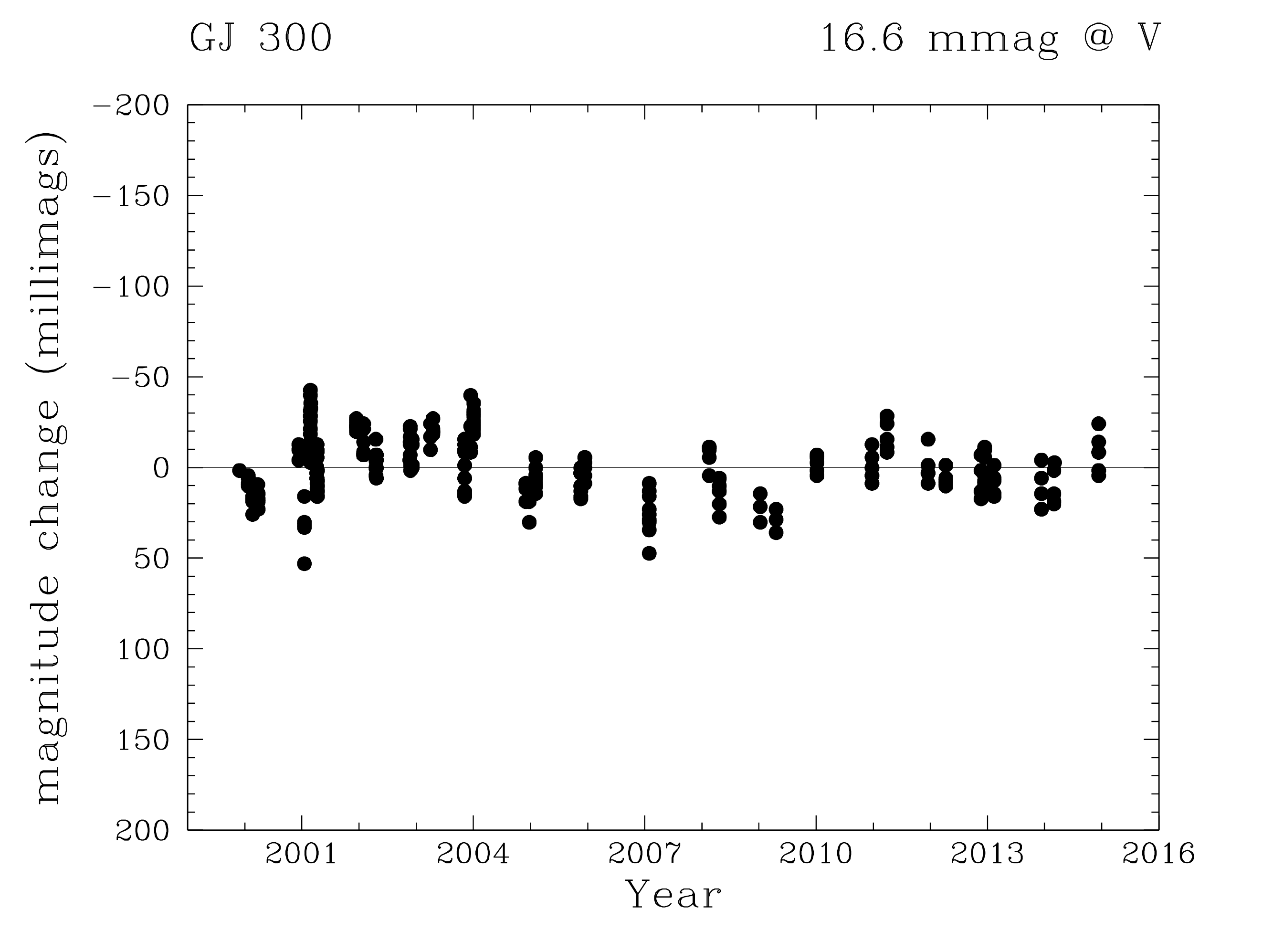}}
\hspace{0.20in}
{\includegraphics[scale=0.32,angle=270]{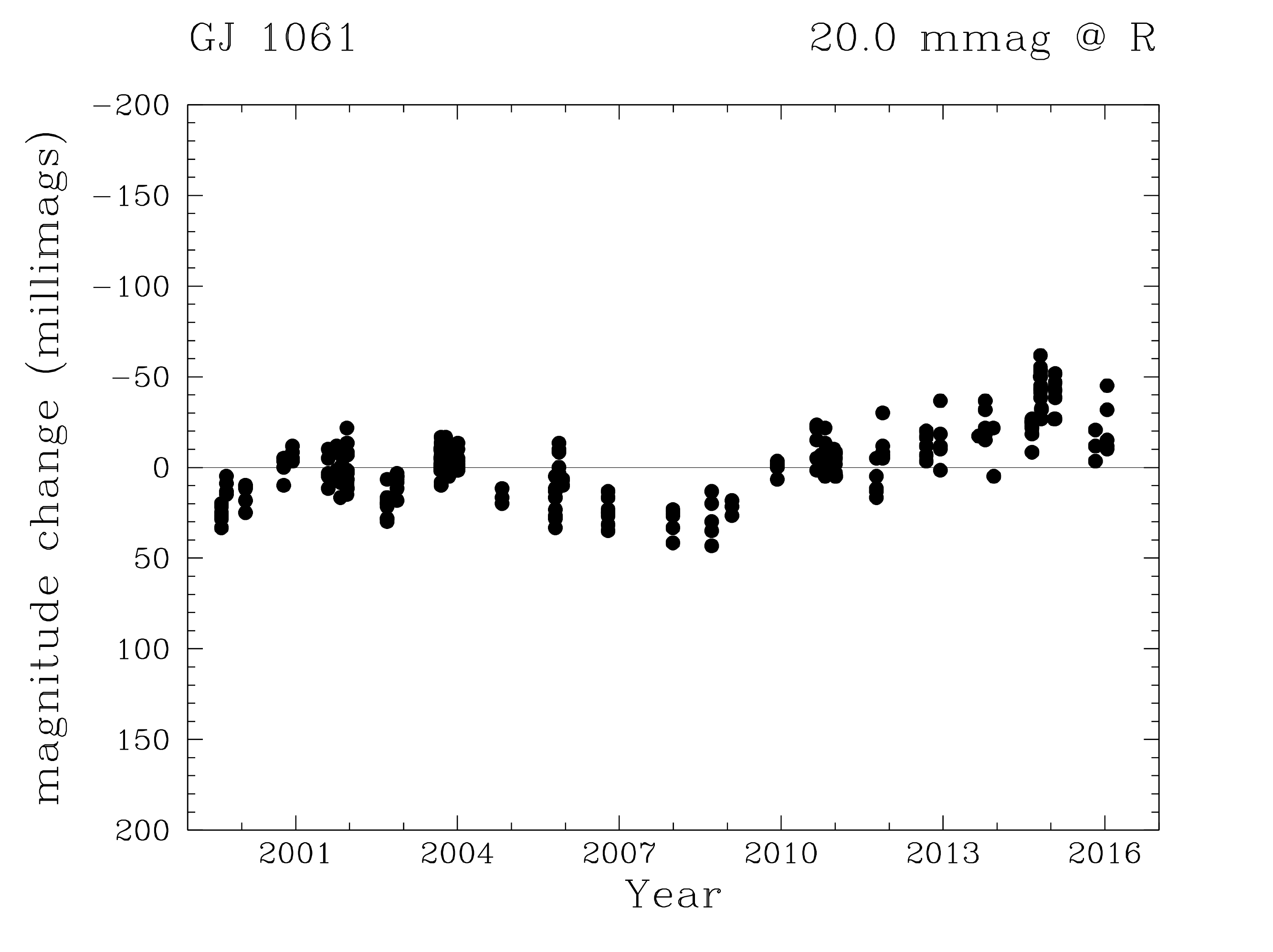}}
\hspace{-0.16in}

\vskip-10pt

{\includegraphics[scale=0.32,angle=270]{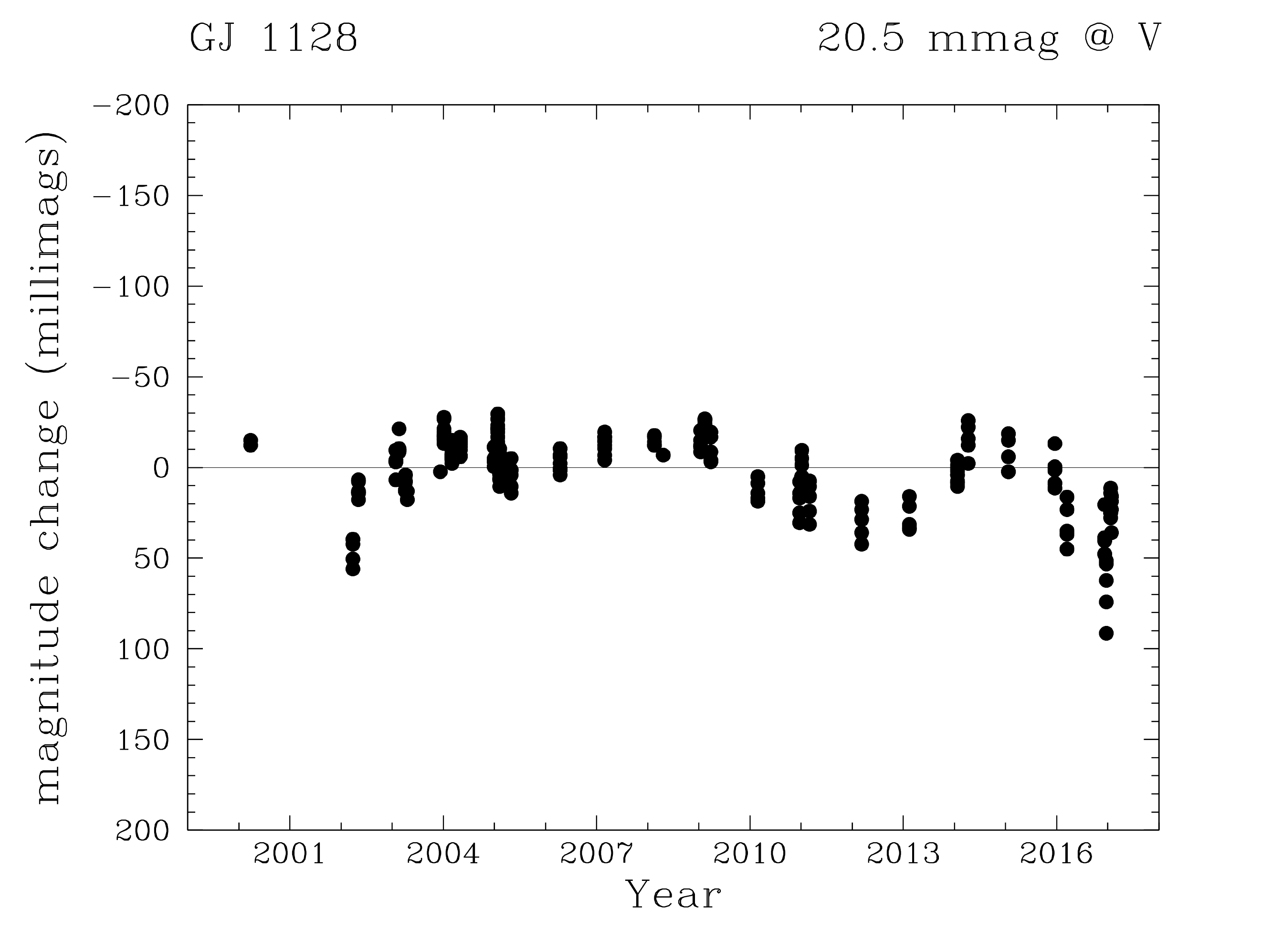}}
\hspace{0.20in}
{\includegraphics[scale=0.32,angle=270]{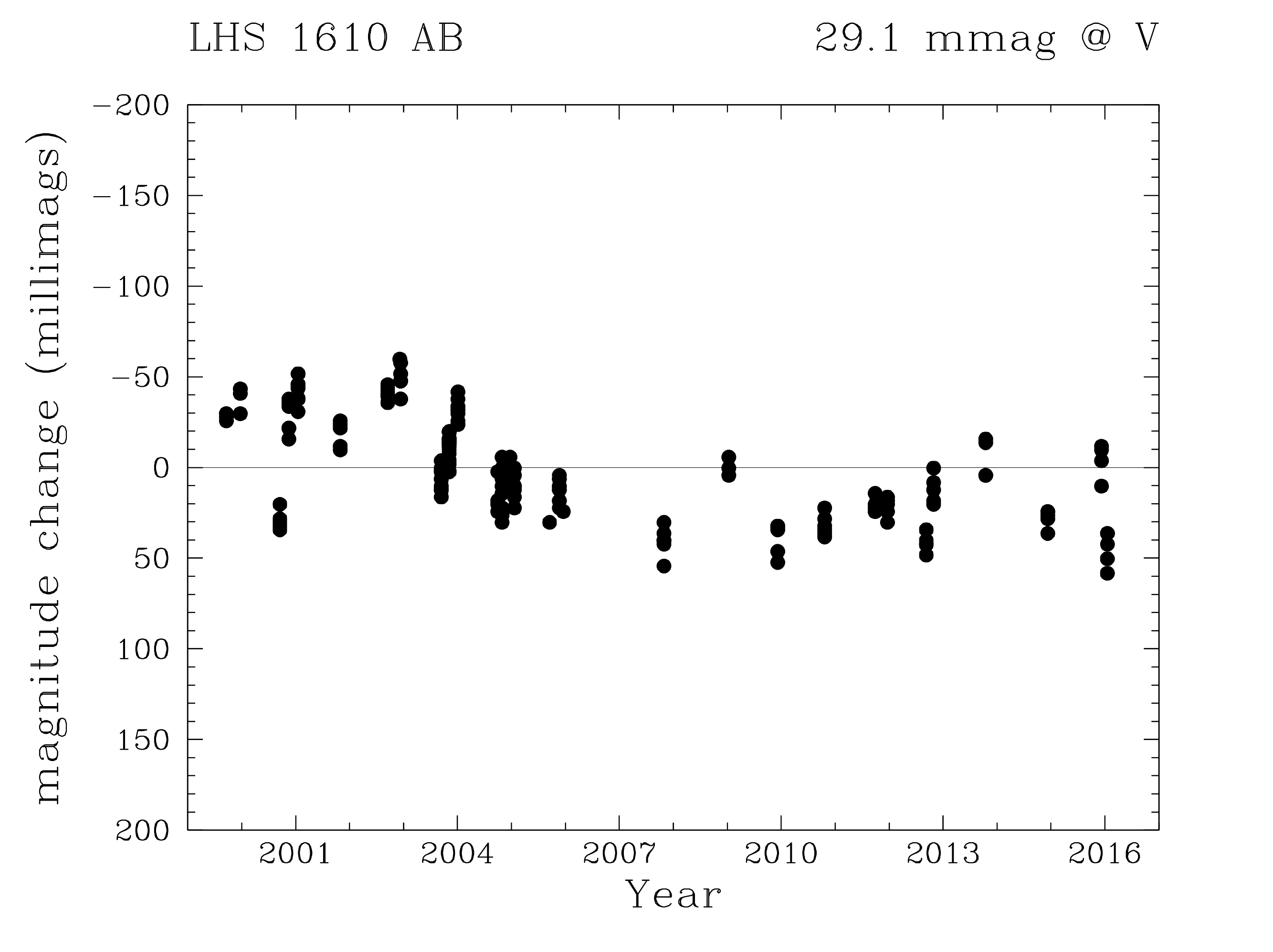}}
\hspace{-0.16in}

\vskip-10pt

{\includegraphics[scale=0.32,angle=270]{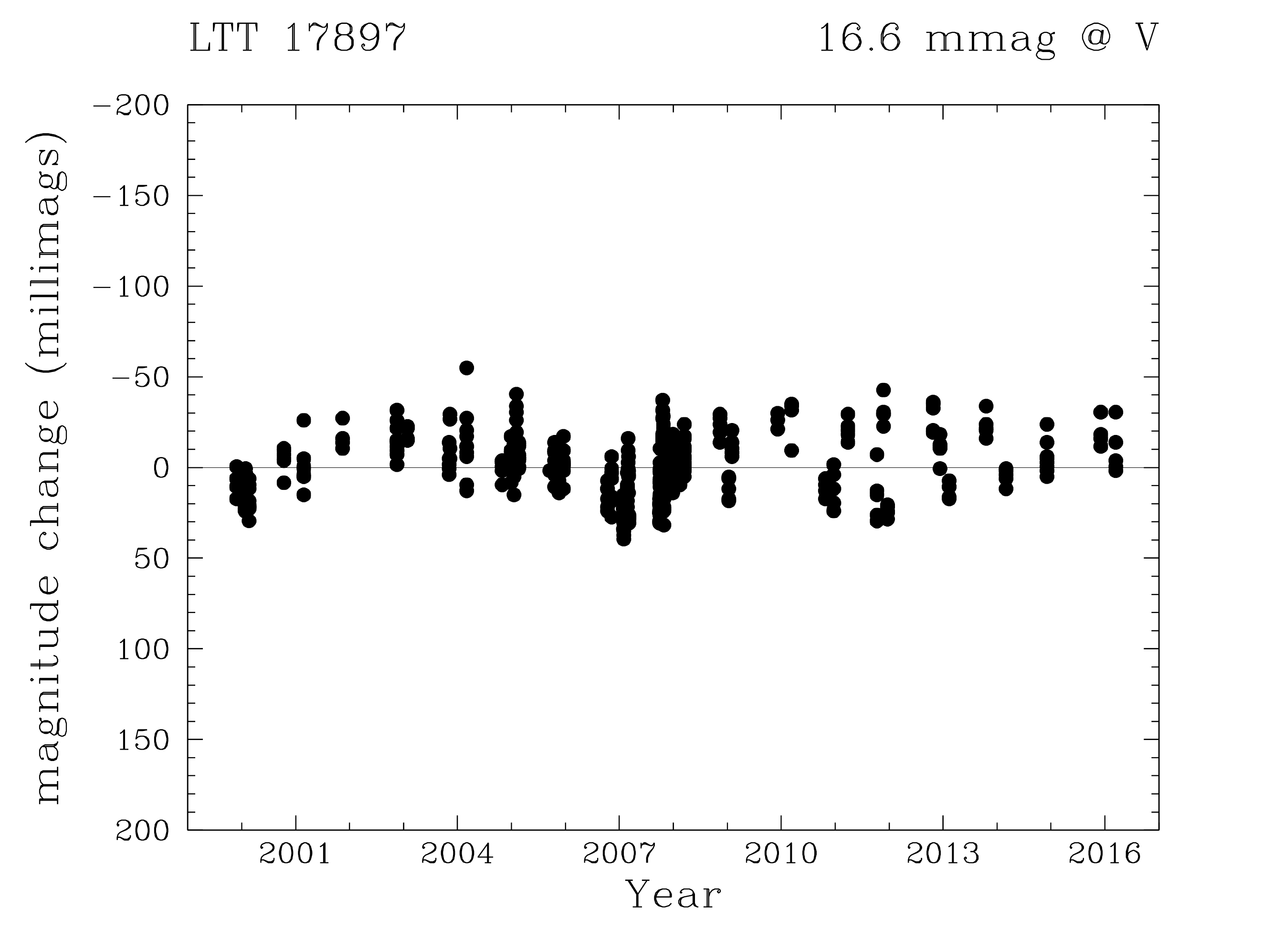}}
\hspace{0.20in}
{\includegraphics[scale=0.32,angle=270] {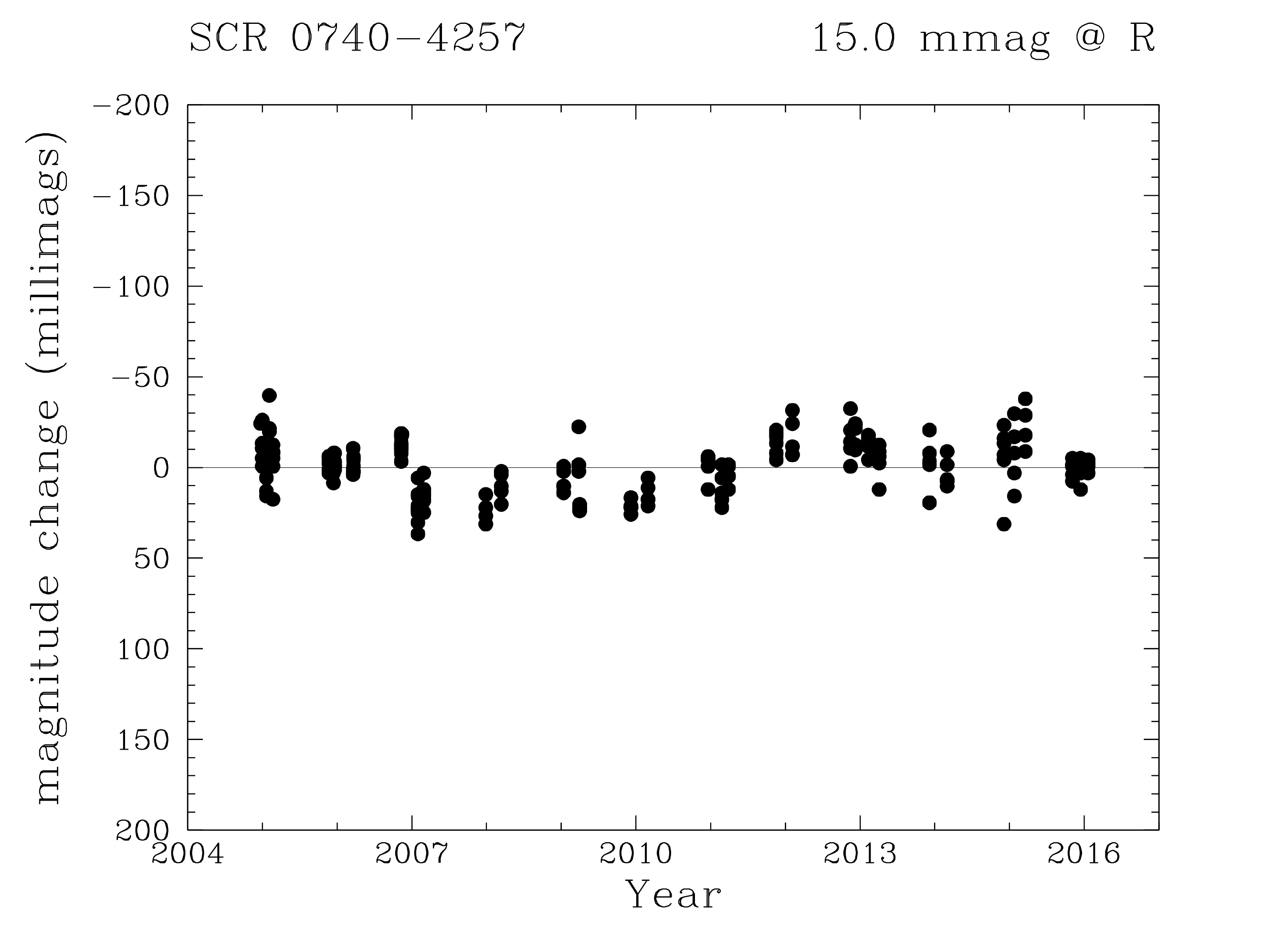}}

\caption{\scriptsize Photometric variability levels are shown for six
  stars with cycles lasting 7--15 years observed at the 0.9m.  Points
  represent brightnesses in single frames and are shown oriented so
  that a star moves up on the plot when it grows brighter.  Star
  names, variability measurements, and the filters used for the
  observations are given at the top of each panel.  Complete cycles
  are evident for all stars except LHS 1610 AB, which may have been
  overluminous from 2000-2004 followed by a complete cycle from
  2004--2014. \label{fig:cycle}}

\end{figure}

$\bullet$ {\it 0335$-$4430} : {\bf GJ 1061} was found to be the 20th
nearest stellar system in \citet{Henry1997}, with $\pi_{trig}$ = 273.4
$\pm$ 5.2 mas determined using photographic plate data from Ianna's
astrometry program in Siding Spring, Australia.  Using the CCD camera
on the 0.9m, in \citet{Henry2006} we reported a parallax of 271.92
$\pm$ 1.34 using 6.3 years of data, and we update that value here to
270.53 $\pm$ 1.18 mas, with a dataset now spanning 16.4 years.  GJ
1061 currently ranks as the 21st nearest system, given the addition of
the closer brown dwarf system WIS 1049-5319AB, but remains the 20th
nearest stellar system and the nearest star discovered since YPC and
{\it Hipparcos}.  The star shows clear long-term variability in the
$V$ band indicative of a stellar cycle, as shown in Figure
\ref{fig:cycle}.

$\bullet$ {\it 0339$-$3525} : {\bf LP 944-020} has been a member of
the 10 pc sample since the first parallax determination of 201.4 $\pm$
4.2 mas by \citet{Tinney1996}, based on less than two years of data.
Our published value of 155.89 $\pm$ 1.03 mas \citep{Dieterich2014} is
from 9.0 years of astrometric data and differs by 11$\sigma$ from the
earlier value.  We update the value here to 156.33 $\pm$ 0.78 mas
using 14.0 years of data.  Although LP 944-020 is not a member of the
5 pc sample, it remains well within 10 pc.

$\bullet$ {\it 0352$+$1701} : {\bf LHS 1610 AB} had a parallax of 70.0
$\pm$ 13.8 mas in the YPC, and we placed the system within 10 pc with
a value of 101.57 $\pm$ 2.07 mas in \citet{Henry2006}.
\citet{Bonfils2013} reported the star to be an SB2, but the detailed
study by \citet{Winters2018} indicates the system to be an SB1 with
orbital period of 10.6 days and that the secondary is likely a brown
dwarf.  Our photometric distance estimate of 9.5 $\pm$ 1.5 pc is
consistent with the trigonometric distance of 9.9 pc, indicating that
the companion contributes minimal flux to the system.  The system
shows a change in brightness in the $V$ band of $\sim$10\% with period
of about a decade, as shown in Figure \ref{fig:cycle}.

$\bullet$ {\it 0504$+$1103} : {\bf LTT 17736 (G 097-015)} was
estimated to be at a distance of 11.5 pc photometrically by
\citet{Weis1986}.  We measure a relative $\pi_{trig}$ = 99.49 $\pm$
0.90 mas, but the reference field is reddened with a correction of
4.07 $\pm$ 0.42 mas.  We adopt a generic correction of 1.50 $\pm$ 0.50
mas, yielding an absolute $\pi_{trig}$ = 100.99 $\pm$ 1.03 mas,
resulting in a distance just inside the 10 pc horizon.

$\bullet$ {\it 0528$+$0938} : {\bf GJ 203} was a member of the 10 pc
sample after results from YPC and {\it Hipparcos}, which resulted in a
weighted mean of $\pi_{trig}$ = 102.60 $\pm$ 2.09 mas.  However,
\citet{Dittmann2014} published $\pi_{trig}$ = 92.9 $\pm$ 2.5 mas,
reducing the weighted mean to 98.61 $\pm$ 1.60 mas.  Our value of
$\pi_{trig}$ = 104.76 $\pm$ 1.52 mas brings the star back into the 10
pc sample with a weighted mean of $\pi_{trig}$ = 101.85 $\pm$ 1.10
mas.

\begin{figure}[ht!]

{\includegraphics[scale=0.32,angle=270]{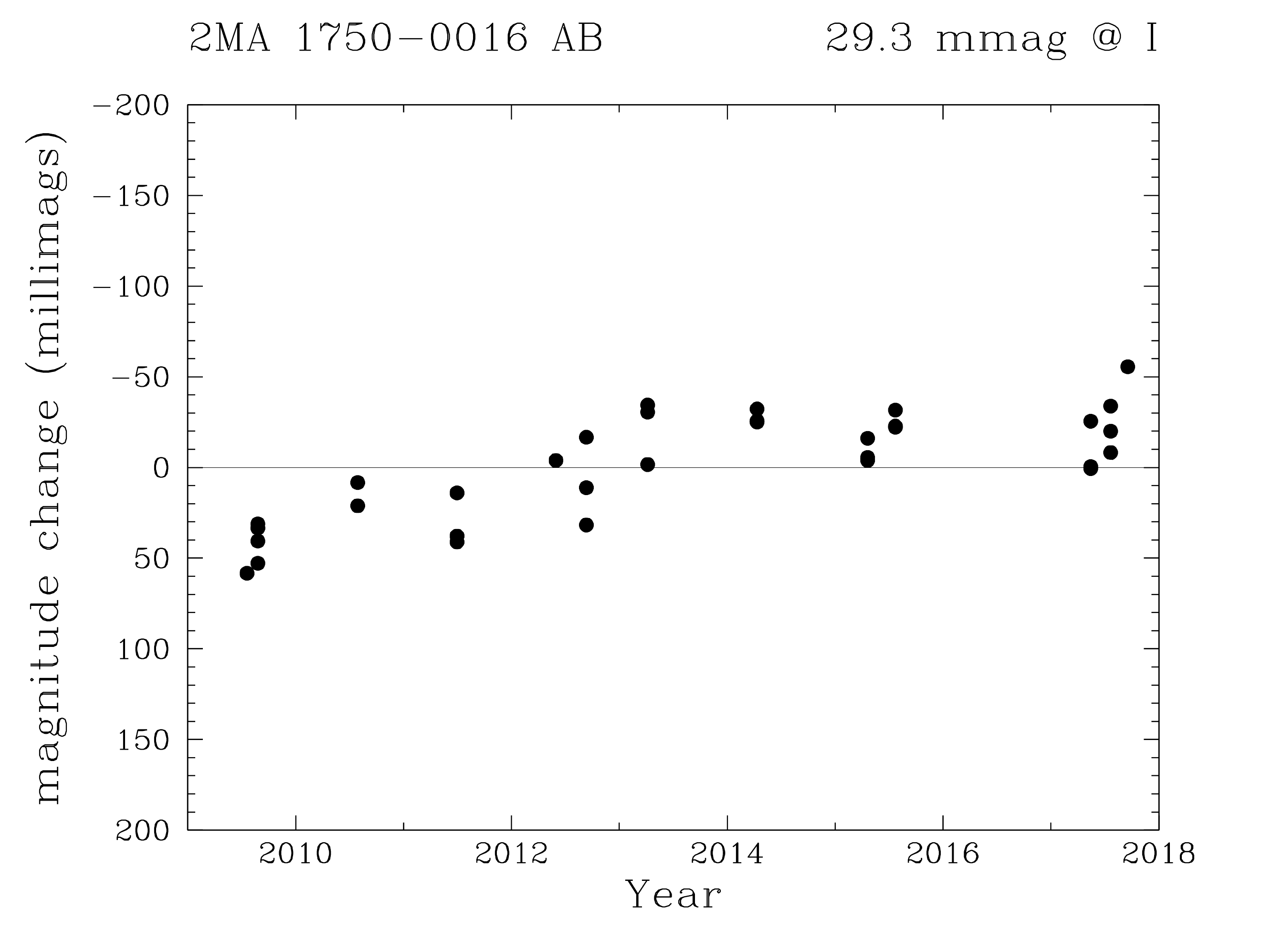}}
\hspace{0.20in}
{\includegraphics[scale=0.32,angle=270]{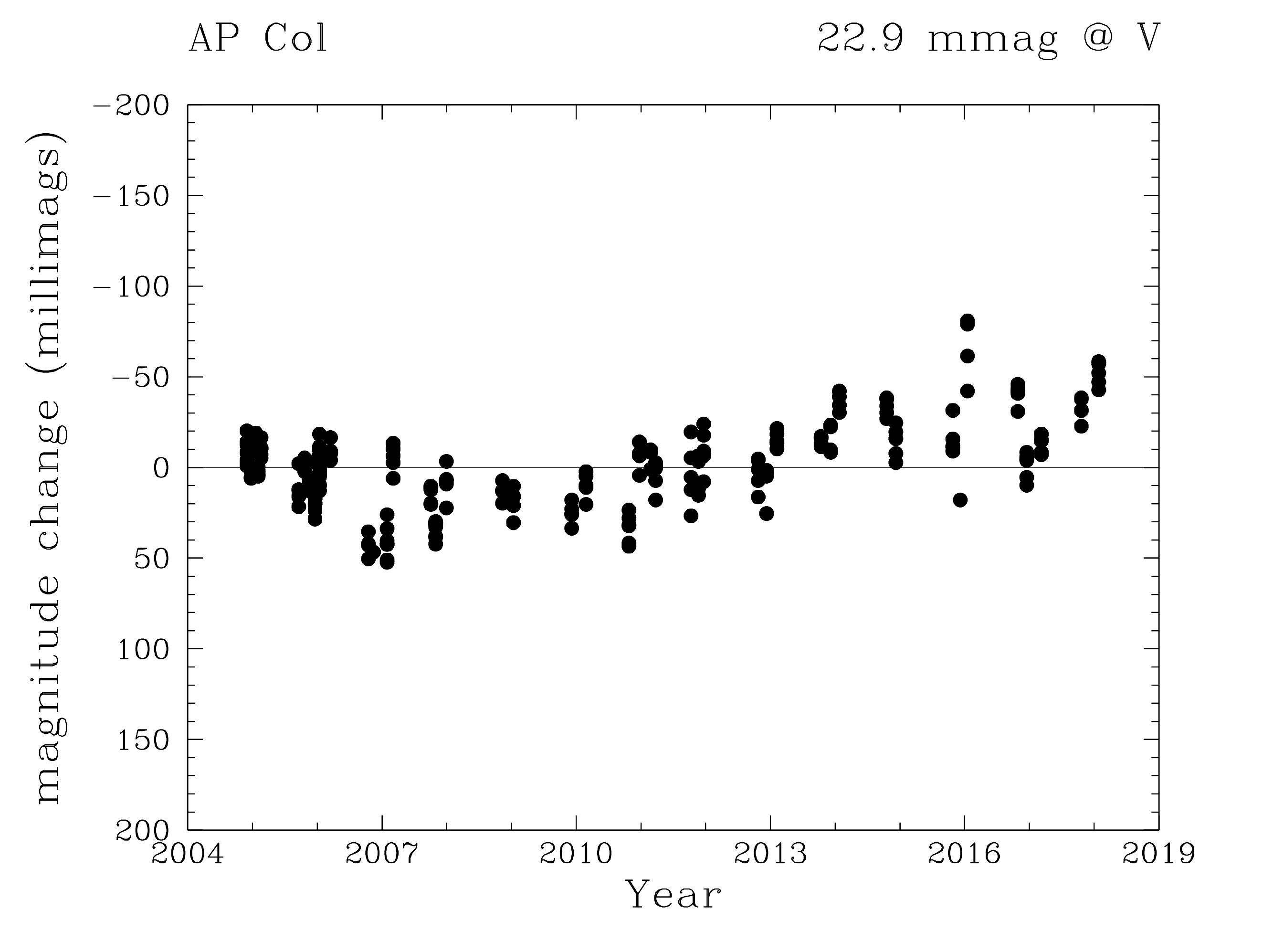}}
\hspace{-0.16in}

\vskip-10pt

{\includegraphics[scale=0.32,angle=270]{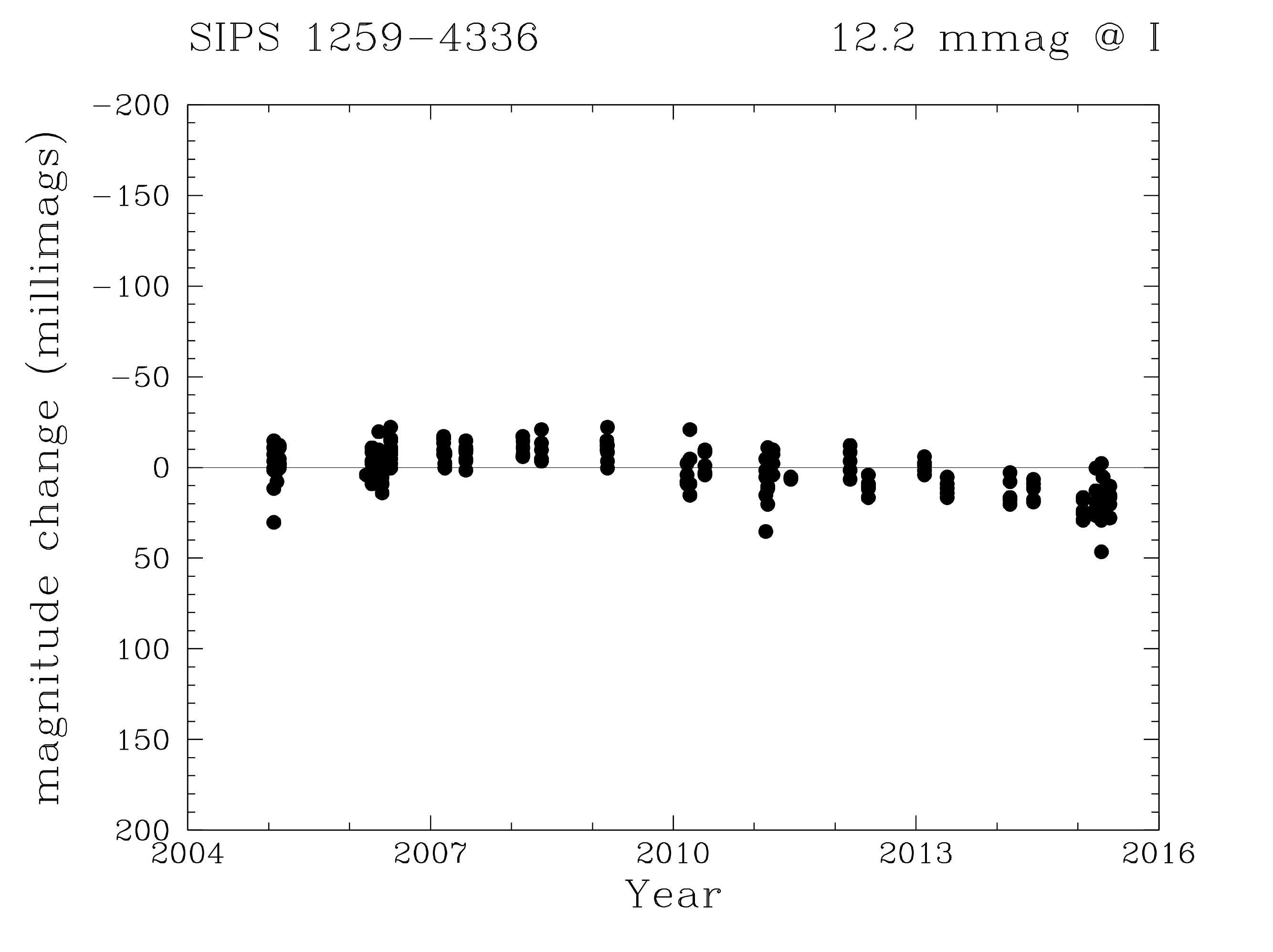}}
\hspace{0.20in}
{\includegraphics[scale=0.32,angle=270]{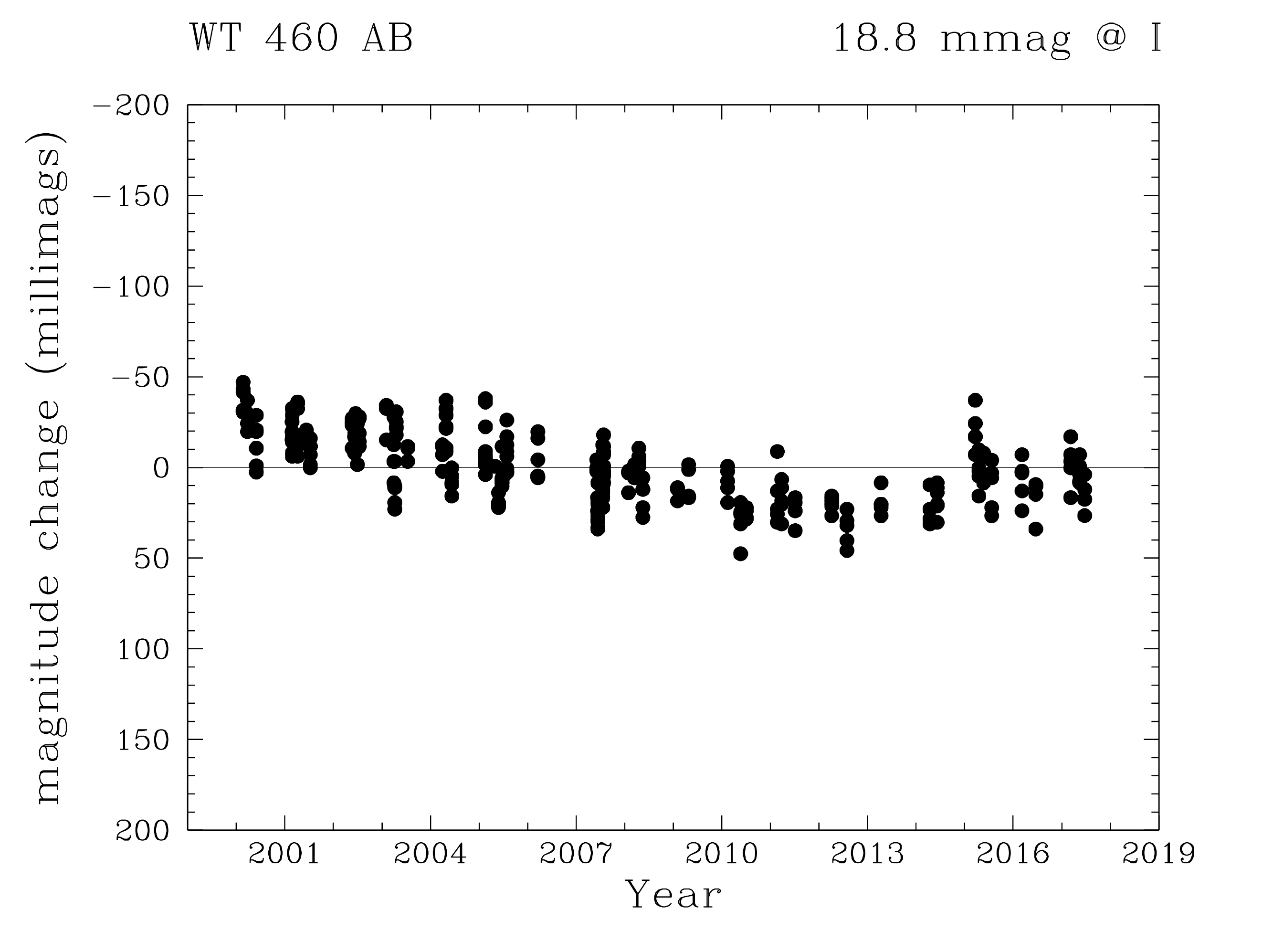}}
\hspace{-0.16in}

\vskip-10pt

{\includegraphics[scale=0.32,angle=270]{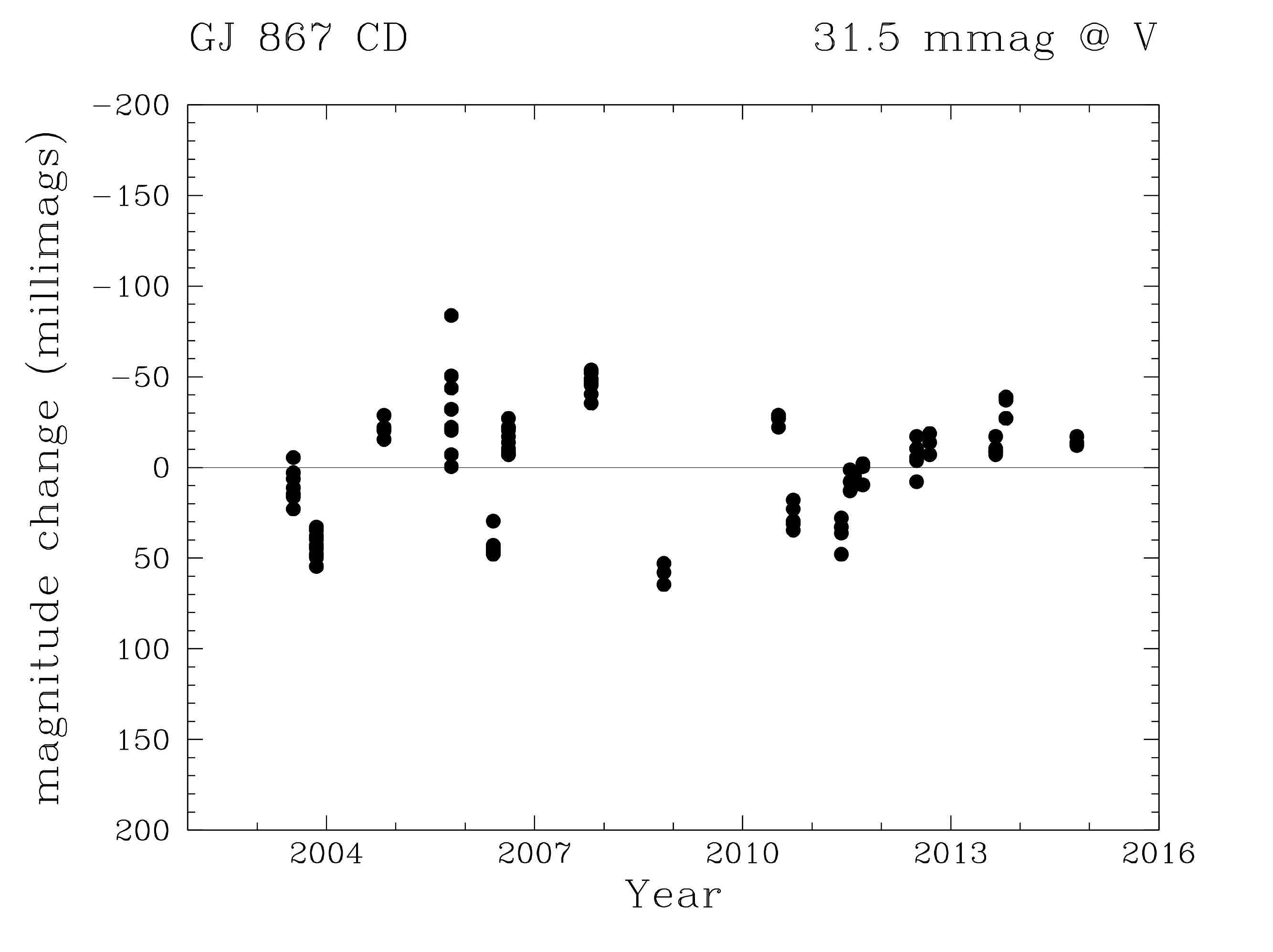}}
\hspace{0.20in}
{\includegraphics[scale=0.32,angle=270]{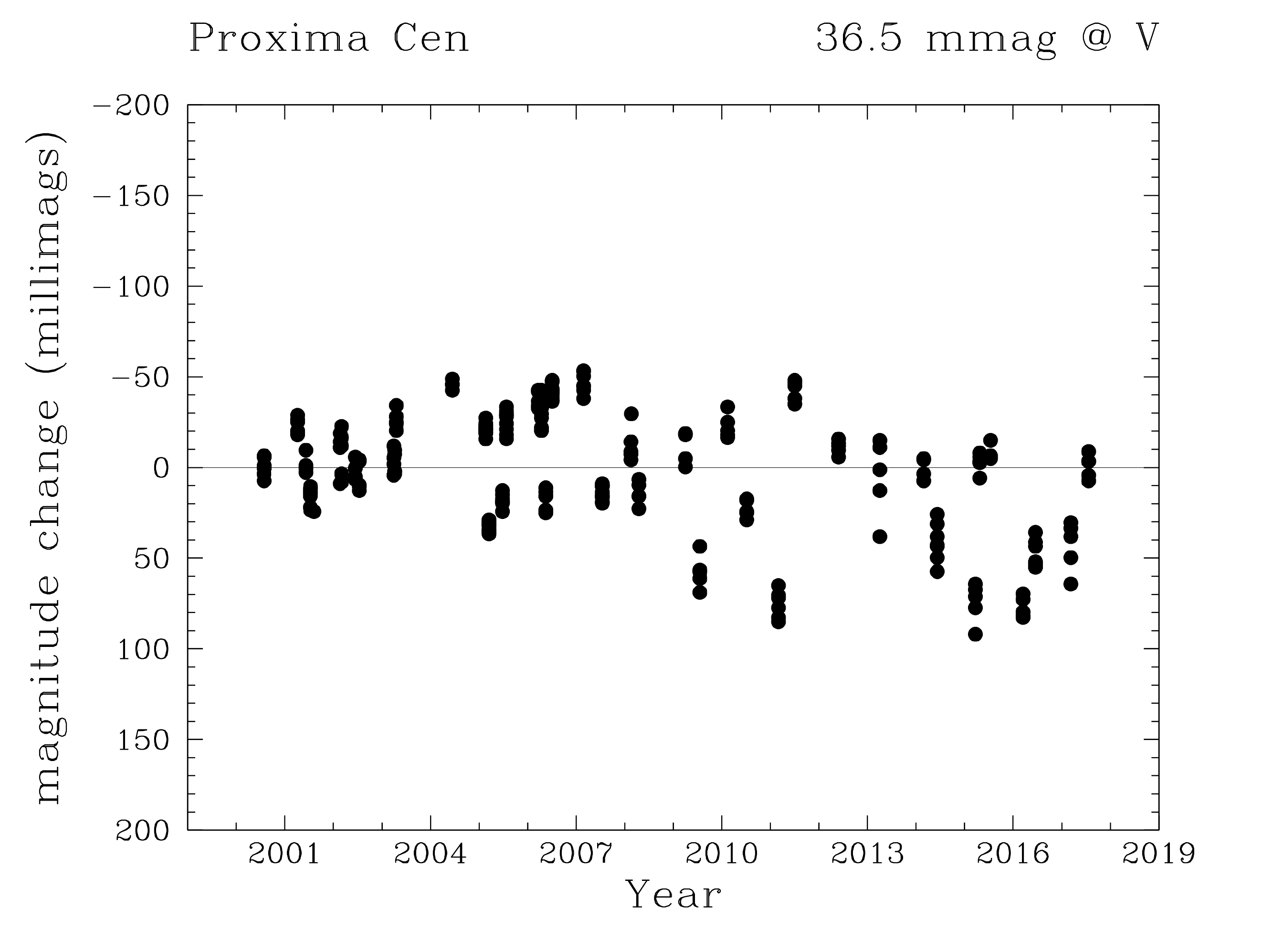}}

\caption{\scriptsize In the top four panels, photometric variability
  levels from 0.9m data are shown for stars with trends in brightness
  that are likely portions of cycles longer than the durations of the
  current datasets.  Points represent brightnesses in single frames
  and are shown oriented so that a star moves up on the plot when it
  grows brighter.  Star names, variability measurements, and the
  filters used for the observations are given at the top of each
  panel.  The cycles are longer than 8 years for 2MA 1750-0016 AB, 13
  years for AP Col, 10 years for SIPS 1259-4336, and 17 years for WT
  460 AB.  The bottom two panels show the two most extreme variability
  measurements among the 75 systems in this paper.  These stars appear
  to be highly spotted, with large, rapid changes in brightness.  GJ
  867 CD is a close binary with an orbital period of 1.8 days where
  components incite activity in one another.  The closest star,
  Proxima Centauri, shows variations of more than 10\% in $V$ during
  the 17 years of coverage. \label{fig:trend}}

\end{figure}
\begin{figure}[ht!]

{\includegraphics[scale=0.70,angle=270]{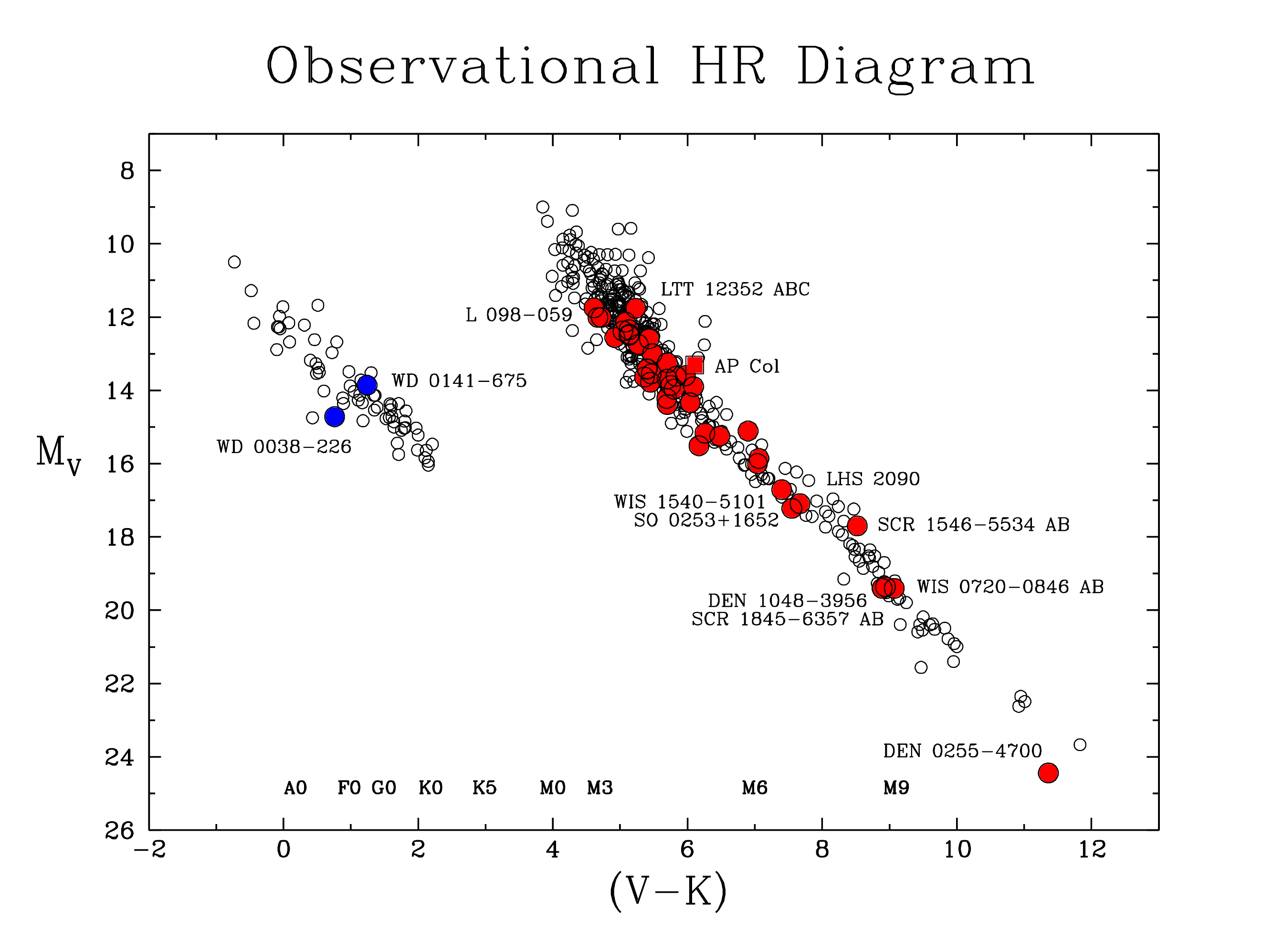}}
\hspace{-0.16in}

\vskip-00pt

\caption{\scriptsize An observational HR diagram is shown for systems
  within 25 pc determined via trigonometric parallaxes from the
  CTIO/SMARTS 0.9m.  Included are the 42 red and brown dwarf systems
  (red points) and two white dwarfs found within 10 pc by RECONS (blue
  points).  The single solid square point represents the pre-main
  sequence star AP Col, which is only 8.6 pc away.  Open black circles
  are additional measurements of targets within 25 pc.  The two
  brightest systems are labeled, as well as the eight intrinsically
  faintest, reddest systems.  \label{fig:hrdia}}

\end{figure}

$\bullet$ {\it 0604$-$3433} : {\bf AP COL} was observed by RECONS
starting in 2004 based upon our photometric distance estimate placing
it at 4.6 pc.  Soon thereafter, \citet{Scholz2005} placed the star (as
LP 949-015) at a distance of 6.1 pc and noted it to be an active M
dwarf.  In 2011, we reported $\pi_{trig}$ = 119.21 $\pm$ 0.98 mas
using 6.5 years of data, placing the M4.5 star at a distance of 8.4
pc.  The star's overluminosity on the HR diagram of Figure
\ref{fig:hrdia} is attributed to youth, supported by its photometric
variability, lithium equivalent width, and gravity indicators
\citep{Riedel2011}.  It is therefore the nearest pre-main sequence
star, and kinematics indicate that it is a likely member of the
$\sim$40 Myr Argus/IC 2391 Association.  We revise the parallax to
116.41 $\pm$ 0.81 mas (8.6 pc) here, and the 13.2 years of data also
reveal a long term change in brightness shown in Figure
\ref{fig:trend} indicative of a cycle lasting longer than the current
dataset.

$\bullet$ {\it 0630$-$7643} : {\bf SCR 0630-7643 AB} is a new binary
discovered during our SCR search \citep{Hambly2004} and first
estimated photometrically to lie at 7.0 pc \citep{Henry2004}.  The
first $\pi_{trig}$ = 114.16 $\pm$ 1.85 mas from \citet{Henry2006} was
based on only 2.0 yr of data, and is virtually unchanged here at
$\pi_{trig}$ = 114.11 $\pm$ 1.44 mas using 13.1 yr of data.
Significant orbital motion is seen, with the B component moving
relative to A from 1\farcs97 at 22$^{\circ}$ on UT 2004 December 23 to
2\farcs66 at 78$^{\circ}$ on UT 2015 December 16, implying an orbital
period on the order of 60 years.  The astrometric reduction was done
using frames with seeing greater than $\sim$1\farcs2, when the two
components merged into a single source for centroiding.

$\bullet$ {\it 0717$-$0501} : {\bf SCR 0717-0501} was first reported
in \citep{Subasavage2005} as a new high proper motion star at an
estimated distance of 15.9 pc based on SuperCOSMOS plate magnitudes
combined with 2MASS.  \citet{Finch2016} reported a parallax of 104.60
$\pm$ 7.90 mas, but our value here of 92.86 $\pm$ 1.61 mas (a slight
update to the value in Winters et al.~2017), places the system beyond
10 pc.

$\bullet$ {\it 0720$-$0846} : {\bf WISE 0720-0846 AB} was revealed to
be a likely nearby star by \citet{Scholz2014} at an estimated distance
of 7.0 pc via a parallax of 142 $\pm$ 38 mas.  \citet{Burgasser2015}
provided a second estimate of the distance via $\pi_{trig}$ = 166
$\pm$ 28 mas and evidence that the system has a brown dwarf companion.
We confirm an unseen secondary that causes the large perturbation
shown in Figure \ref{fig:perts}; once removed, we measure measure
$\pi_{trig}$ = 148.80 $\pm$ 1.08 mas (6.7 pc) using 4.0 yr of data.
With the relatively short time coverage, we can only say that the
period is long, but the similar trigonometric and photocentric
distances indicate that the secondary contributes minimal light to the
system, consistent with it being of lower mass than the M9.0V primary,
i.e., likely a brown dwarf.  With $\mu$ = 109 mas/yr, this is one of
four new systems reported here with proper motions less than 200
mas/yr, ranking as the second slowest moving system known within 10
pc.  The star has been identified as possibly disrupting the Sun's
Oort Cloud $\sim$70,000 years ago \citet{Mamajek2015}.


$\bullet$ {\it 0736$+$0704} : {\bf LTT 17993 AB (G 089-032 AB)} was
first reported to be a close double by \cite{Henry1999} with a
separation of 0\farcs7 using infrared speckle imaging.  The system was
reported to have $\pi_{trig}$ = 116.60 $\pm$ 0.97 mas using 6.1 yr of
data in \citet{Henry2006}, and no perturbation was detected at that
time.  FWHM measurements of the system at the 0.9m from 2000-2010 were
always $\sim$1\farcs5, but steadily dropped to 1\farcs0 by 2018,
implying a decreasing separation, and a long-term perturbation is now
seen.  A tentative orbital fit to residuals on both RA and DEC axes
implies a period longer than the 18.0 years of available data, and
that signal has been removed for the parallax of 117.59 $\pm$ 0.83 mas
in Table 1.  This is a promising system for accurate mass
determinations.

$\bullet$ {\it 0740$-$4257} : {\bf SCR 0740-4257} was first reported
in \citep{Subasavage2005}, in which we estimated the distance to be
10.0 pc using SuperCOSMOS $BRI$ plate magnitudes combined with 2MASS
$JHK$.  We then obtained accurate $VRI$ magnitudes, and when combined
with 2MASS estimated a distance of only 7.2 pc in \citet{Winters2011}.
Here we present the first parallax, $\pi_{trig}$ = 127.71 $\pm$ 0.76
mas, based on 11.1 years of data, placing the system at 7.8 pc.  This
is supported by a URAT value of $\pi_{trig}$ = 122.7 $\pm$ 4.0 mas.
As shown in Figure \ref{fig:cycle}, the star appears to show a
low-level photometric variation in $R$ with period of $\sim$7 years.

$\bullet$ {\it 0751$-$0000} : {\bf GJ 1103 (LHS 1951)} was reported in
\citet{Luyten1979} to have a companion 3\arcsec~from the M4.5V star.
Our own monitoring from the 0.9m from 2004-2018 in the $V$ filter does
not reveal any companion at that separation, nor to separations as
small as the seeing, 0\farcs9 in the best frames.  In addition, we
find no perturbation on the position of the photocenter after solving
for parallax and proper motion over 13.1 years of monitoring.
Blinking of plate images reveals a background star as the ``B''
component at that separation and position angle in POSS-1 plates from
1953, so we conclude that the reported ``companion'' is a background
star.


$\bullet$ {\it 0815$-$2344} : {\bf UPM 0815-2344 AB} is the first
system from the USNO Proper Motion (UPM) survey by \citet{Finch2014}
confirmed to be within 10 pc via a RECONS parallax measurement,
$\pi_{trig}$ = 105.05 $\pm$ 0.87 mas (9.5 pc) reported here.  This
distance is consistent with their photometric distance estimate of
11.1 $\pm$ 2.6 pc, and is confirmed by a URAT parallax of 102.9 $\pm$
3.4 mas.  The B component creates a slight tail in the contour plot,
about 1\farcs5 from the primary; no orbital motion nor perturbation is
yet seen in the 3.2 years of data.  With $\mu$ = 138 mas/yr, this is
one of four new systems reported here with proper motions less than
200 mas/yr, ranking fifth among all systems within 10 pc.

$\bullet$ {\it 0818$-$6818} : {\bf L 098-059} slips into the 10 pc
sample with $\pi_{trig}$ = 101.81 $\pm$ 2.37 mas.  The error is large
because of a relatively faint field of reference stars compared to the
$R$ = 10.61 star.

$\bullet$ {\it 0835$-$0819} : {\bf 2MA 0835-0819} has a $\pi_{trig}$ =
117.3 $\pm$ 11.2 mas (8.5 pc) in \citet{Andrei2011}, but the error was
too large for inclusion in the 10 pc sample.  Parallaxes by
\citet{Weinberger2016} (137.49 $\pm$ 0.39), \citet{Dahn2017} (138.71
$\pm$ 0.58 mas), and our value here (138.33 $\pm$ 0.93 mas), are all
virtually identical and place this brown dwarf within 10 pc, and
substantially closer (7.2 pc) than the original distance.

$\bullet$ {\it 0858$+$0828} : {\bf LTT 12352 ABC (G 041-014 ABC)} is a
close triple in which all three components are separated by less than
1$\arcsec$, as discussed in \cite{Henry2006}.  The five reference
stars are faint, and with an average integration time of only 37
seconds the astrometric residuals are relatively high.  Nonetheless,
even after 18.0 years, no perturbation is seen, and images have FWHM
as small as 1\farcs0 with no obvious elongation.


$\bullet$ {\it 0915$-$1035} : {\bf LHS 6167 AB} was identified as a
possible nearby star by \citet{Finch2016} with $\pi_{trig}$ = 134.9
$\pm$ 12.1 mas, although the large parallax error precluded its entry
to the 10 pc sample.  A parallax of 103.30 $\pm$ 1.00 mas using 9.3 yr
of data was reported in \citet{Bartlett2017}, adding the system to the
sample.  The binary has been resolved several times, always at a
separation less than 0\farcs2 (see Bartlett et al.~2017, Table 9).  We
provide an updated parallax value of $\pi_{trig}$ = 103.54 $\pm$ 0.77
mas based on 13.3 yr of data that includes the orbital fit (see Table
2) with period of 4.8 yr shown in Figure \ref{fig:orbit}.  The system
flared by 0.25 mag in the $V$ band on UT 2013 April 02.

$\bullet$ {\it 1048$-$3956} : {\bf DEN 1048-3956} was targeted by
RECONS in both our 0.9m and 1.5m astrometry programs, yielding
parallaxes of 247.71 $\pm$ 1.55 mas \citep{Jao2005} and 249.78 $\pm$
1.81 mas \citep{Costa2005}, respectively.  The 0.9m result was updated
in \citet{Lurie2014} to $\pi_{trig}$ = 248.08 $\pm$ 0.61 mas, where
orbiting planets with masses of 1--2 Jupiters were eliminated for
periods of 2--12 years.  Now with 14.9 years of data, we adjust the
parallax slightly to $\pi_{trig}$ = 247.78 $\pm$ 0.60 mas, which now
ranks this M8.5V star as the 29th nearest system to the Sun.

%

$\bullet$ {\it 1141$-$3624} : {\bf SIPS 1141-3624} was first reported
in \citet{Deacon2005a} as part of a survey for new high proper motion
stars using a combination of data from the SuperCOSMOS sky survey and
2MASS.  We estimate the distance photometrically to be 9.9 pc, whereas
our $\pi_{trig}$ = 116.12 $\pm$ 1.30 mas is the first, corresponding
to 8.6 pc.  A URAT value of $\pi_{trig}$ = 107.1 $\pm$ 4.7 mas
confirms the systems's proximity.  The star appears to exhibit a low
amplitude photometric cycle in the $R$ band of duration $\sim$7 yr.

$\bullet$ {\it 1155$-$3727} : {\bf 2MA 1155-3727} is an object near
the stellar/substellar border reported to have $\pi_{trig}$ = 104.40
$\pm$ 4.70 mas in \citet{Faherty2012}.  Values from both
\citet{Weinberger2016} ($\pi_{trig}$ = 84.36 $\pm$ 0.85 mas) and our
value here ($\pi_{trig}$ = 82.76 $\pm$ 1.74 mas) place this object
well beyond 10 pc.

$\bullet$ {\it 1159$-$5247} : {\bf 1RXS 1159-5247} was put on the 0.9m
program in 2009 given our photometric distance estimate of 10.6 pc.
\citet{Sahlmann2014} reported a parallax of $\pi_{trig}$ = 105.54
$\pm$ 0.12 mas for the star, which was part of an astrometric search
for planets around 20 nearby stars.  We measure $\pi_{trig}$ = 106.25
$\pm$ 0.69 mas, consistent with the \citet{Sahlmann2014} value.

$\bullet$ {\it 1214$+$0037} : {\bf GJ 1154} was reported to be an
unresolved SB2 by \citet{Bonfils2013}, with variable spectral-line
width.  We detect no perturbation in 4.8 years of data, and the value
of $\pi_{trig}$ = 124.68 $\pm$ 1.83 mas yields a distance of 8.0 $\pm$
0.1 pc that is marginally consistent with the photometric distance of
6.3 $\pm$ 1.0 pc.  Although the distance offset may indicate flux
contributed by an unseen secondary, we await confirmation of the
companion before listing the system as a close double.

$\bullet$ {\it 1259$-$4336} : {\bf SIPS 1259-4336} is a high proper
motion star ($\mu$ = 1.145\arcsec/yr) first reported in
\citet{Deacon2005a} to be at a distance of only 3.6 pc based on
$\pi_{trig}$ = 276 $\pm$ 41 mas.  We estimate the distance to be 8.1
pc photometrically, and determine a distance of 7.8 pc from
$\pi_{trig}$ = 127.93 $\pm$ 0.50 mas, more than twice as distant as
the initial parallax estimate.  The low formal error is representative
of our very best parallaxes.  A slight long-term trend in brightness
over a decade in the $I$ band is evident, as shown in Figure
\ref{fig:trend}.

$\bullet$ {\it 1411$-$4132} : {\bf WT 460 AB} entered the 10 pc sample
after our first determination of $\pi_{trig}$ = 107.41 $\pm$ 1.52 mas
\citep{Henry2006}.  A low mass companion was first reported by
\citet{Montagnier2006} at a separation of 0\farcs51 mas and brightness
difference in the $H$ band of 2.47 mag.  The companion causes a large
perturbation shown in Figure \ref{fig:perts} that we remove to derive
an updated parallax of 108.53 $\pm$ 0.74 mas using 17.3 years of data.
The photocentric shifts are 250 mas in both RA and DEC and show no
sign of abating, implying a period of at least several decades.  As
shown in Figure \ref{fig:trend}, this system has been found to exhibit
a long term photometric trend in the $I$ band, with the flux dropping
$\sim$5\% from 2000--2012 followed by a turnaround in the past few
years.

$\bullet$ {\it 1416$+$1348} : {\bf SDSS 1416+1348 AB} is a brown dwarf
pair separated by 9\arcsec~for which the first parallax of 109.7 $\pm$
1.3 mas was published by \citep{Dupuy2012}.  Our value of 106.77 $\pm$
1.24 mas derived after 7.0 years of data is consistent, and secures
this double brown dwarf as one of only four known within 10 pc.

$\bullet$ {\it 1429$-$6240} : {\bf PROXIMA CENTAURI} We have observed
the nearest star to the Sun for 17.0 years and measure a parallax of
769.66 $\pm$ 0.83 mas, consistent with the weighted mean of 768.74
$\pm$ 0.30 mas from YPC, {\it Hipparcos}, and $HST$
\citep{Benedict1999}.  Proxima is undoubtedly a highly spotted star,
with variations of more than 10\% in the $V$ band, as shown in Figure
\ref{fig:trend}.  In fact, its measured variability of 36.5 mmag is
the largest for any star among the 75 discussed in this paper.  A
rotation period of $\sim$83 days has been derived for Proxima
\citep{Benedict1998} and the star has been known to flare (e.g.,
MacGregor et al.~2018), although we have not observed a major flare
during our observations.  

$\bullet$ {\it 1540$-$5101} : {\bf WISE 1540-5101} is a high proper
motion M dwarf with $\mu$ = 1\farcs981/yr that had two crude
parallaxes of 228 $\pm$ 24 mas \citep{PerezGarrido2014} and 165 $\pm$
41 \citep{Kirkpatrick2014}.  Our photometric distance estimate of 5.1
pc is consistent with our $\pi_{trig}$ 187.93 = $\pm$ 1.15 mas (5.3
pc).  The measurement is based on only 2.0 yr of data, although we
deem the result reliable because we have sampled the extrema of the
parallax ellipse and reductions with earlier frame sets are consistent
with the value presented here.

$\bullet$ {\it 1542$-$1928} : {\bf GJ 595 AB} has long been known to
be nearly 10 pc away, with $\pi_{trig}$ = 99.42 $\pm$ 3.46 mas from
the weighted mean parallaxes of YPC (101.60 = $\pm$ 4.40 mas) and {\it
  Hipparcos} (95.50 = $\pm$ 5.59 mas) values.  At $V=$ 11.84, this is
near the faint limit of {\it Hipparcos}, so the $\pi_{trig}$ error was
relatively large.  \citet{Nidever2002} discovered a companion with
orbital period 63 days and minimum mass 60 $M_{Jup}$ that may have
confounded previous parallax efforts.  We confirm the companion in our
astrometry data using 5.4 years of data, and derive a nearly edge-on
orbit (see Table 2) with a period of 62.0 $\pm$ 0.1 days and
photocentric semimajor axis of 9 mas.  The inclination of 95$^{\circ}$
implies that the companion is a brown dwarf.  After fitting the orbit
shown in Figure \ref{fig:orbit}, we measure $\pi_{trig}$ = 108.02
$\pm$ 1.31 mas, placing the system comfortably within 10 pc.

$\bullet$ {\it 1546$-$5534} : {\bf SCR 1546-5534 AB} was discovered by
the RECONS team \citep{Boyd2011}, where we estimated the distance to
be 6.7 pc using SuperCOSMOS $BRI$ plate magnitudes combined with 2MASS
$JHK$.  Combining accurate $VRI$ \citep{Winters2011} with 2MASS, we
derived an identical photometric distance estimate of 6.7 pc.  With
6.1 years of astrometric data, we now see a clear perturbation shown
in Figure \ref{fig:perts} caused by a companion unseen in our images.
The perturbation is quite large with photocenter shifts of $\sim$100
mas in both RA and DEC, and the orbit has not clearly wrapped.
Removal of the perturbation results in a relative $\pi_{trig}$ =
103.11 $\pm$ 0.87 mas, placing the system near 10 pc.  The reference
stars yield a correction to absolute parallax of 3.42 mas, so we have
adopted a generic correction of 1.50 $\pm$ 0.50 mas.  The offset
between the trigonometric and photocentric distances indicates that
the secondary contributes significant light to the system and is not
of low mass, consistent with the large perturbation.

$\bullet$ {\it 1645$-$1319} : {\bf 2MA 1645-1319} was reported in
\citet{Faherty2012} to have $\pi_{trig}$ = 109.9 $\pm$ 6.1 mas, but
our value of 90.12 $\pm$ 0.82 mas from \cite{Dieterich2014} using 4.0
years of data indicates that this star is not within 10 pc.  We
provide a slightly revised value of 90.89 $\pm$ 0.80 mas based on more
than twice as much data (8.2 years).

$\bullet$ {\it 1731$+$2721} : {\bf 2MA 1731+2721} was reported to have
$\pi_{trig}$ = 113.8 $\pm$ 7.0 mas in \citet{Dittmann2014}, but here
we find $\pi_{trig}$ = 85.54 $\pm$ 1.69 mas based on 6.2 years of
data, placing the star well beyond 10 pc.

$\bullet$ {\it 1737$-$4419} : {\bf GJ 682} has a weighted mean
parallax of 199.65 $\pm$ 2.30 mas from YPC and {\it Hipparcos},
placing it just beyond 5 pc.  During observations spanning 14.0 years,
the target star moves in front of a few background stars, but we deem
our $\pi_{trig}$ = 203.49 $\pm$ 1.30 mas value reliable, indicating
that the star is a new addition to the 5 pc sample.  A generic
correction of 1.50 $\pm$ 0.50 mas from relative to absolute parallax
has been used because the background stars are reddened.

$\bullet$ {\it 1750$-$0016} : {\bf 2MA 1750-0016 AB} is a brown dwarf
reported by \citet{Andrei2011} to have $\pi_{trig}$ = 108.5 $\pm$ 2.6
mas.  \citet{Dahn2017} confirmed its proximity via a parallax of
110.59 $\pm$ 0.52 mas.  We have discovered an unseen companion,
presumably a lower mass brown dwarf, that causes the perturbation on
the photocenter shown in Figure \ref{fig:orbit}.  An orbital fit to
the motion (see Table 2) yields a period of 6.7 years and semimajor
axis of 13 mas, and once removed, we find a parallax of 111.47 $\pm$
1.18 mas (9.0 pc).  This is the fourth binary brown dwarf known within
10 pc and the third closest, along with WISE 1049-5319 AB (2.0 pc),
$\epsilon$ Indi BC (3.6 pc), and SDSS 1416+1348 AB (9.1 pc).  As shown
in Figure \ref{fig:trend}, the system also shows a trend in flux,
growing brighter in the $I$ band by $\sim$6\% over 8 years.

$\bullet$ {\it 1843$-$5436} : {\bf LHS 5341} is a virtually unexplored
single red dwarf for which we measure $\pi_{trig}$ = 101.57 $\pm$ 1.01
mas, placing the star at 9.8 pc.  A URAT value of $\pi_{trig}$ = 105.9
$\pm$ 5.5 mas supports its membership in the 10 pc sample.  This is a
classic case of a nearby star in the southern sky about which almost
nothing is known; the only previous study of this star is our own
indicating that it is not variable by 20 mmag over $\sim$6 years in
the $R$ band \citep{Hosey2015}.

$\bullet$ {\it 1845$-$6357} : {\bf SCR 1845-6357 AB} is the nearest
star discovered via the SuperCOSMOS-RECONS effort.  The primary is an
M8.5V star first reported in \citet{Hambly2004}, with a brown dwarf
secondary first imaged by \citet{Biller2006} at a separation of
1\farcs17.  We published the first parallax estimate in
\citet{Deacon2005b}, $\pi_{trig}$ = 282 $\pm$ 23 mas, using scans of
eight SuperCOSMOS scans of plates in the UK Schmidt Telescope Unit
plate library.  We followed this with a measurement from the 0.9m
program in \citet{Henry2006}, determining $\pi_{trig}$ = 259.45 $\pm$
1.11 mas.  Here we provide an updated value that is significantly
smaller, at 251.27 $\pm$ 0.54 mas.  The parallax has changed because
of the slow orbital pull by the brown dwarf companion that affected
the initial parallax value determined using only 2.6 years of data.
The value presented here uses 14.5 years of data for which the
perturbation shown in Figure \ref{fig:perts} has been fit and removed.
The orbital period is likely a century or more, so the fit is
certainly not the true orbit, but it does improve the parallax
measurement.  At 4.0 pc, the system currently ranks as the 27th
nearest.

$\bullet$ {\it 2049$-$4012} : {\bf SCR 2049-4012 AB} has an estimated
photometric distance 8.0 pc, which is closer than the 9.6 pc result
from the parallax $\pi_{trig}$ = 104.39 $\pm$ 0.87 mas.  URAT provides
an additional measurement of $\pi_{trig}$ = 107.6 $\pm$ 4.4 mas,
confirming that the system is within 10 pc.  The offset between the
photometric and trigonometric distances is likely due to the unseen
companion causing the perturbation with orbital period of $\sim$1.8 yr
shown in Figure \ref{fig:orbit}, although the other orbital elements
(see Table 2) are likely unreliable.  With $\mu$ = 68 mas/yr, this is
one of four new systems reported here with proper motions less than
200 mas/yr, and ranks as the slowest-moving system known within 10 pc.


$\bullet$ {\it 2050$-$3424} : {\bf 2MA 2050-3424} is from the southern
SUPERBLINK survey, reported in \citet{Lepine2008}.  Our photometric
distance estimate of 9.5 pc is consistent with the $\pi_{trig}$ =
105.61 $\pm$ 1.10 mas (9.5 pc) RECONS measurement.  URAT also finds a
value that places the star within 10 pc, $\pi_{trig}$ = 101.0 $\pm$
5.6 mas.

$\bullet$ {\it 2238$-$2036} : {\bf GJ 867 CD} is part of a quadruple
system, comprised of two close doubles, AB and CD,\footnote{Here we
  describe the system as AB-CD rather than AC-BD as used by
  \citet{Davison2014} --- we have reassigned the nomenclature to be
  consistent with the likely order of brightness at $V$.}  separated
by 24\arcsec, corresponding to a projected separation of $\sim$200 AU.
The AB pair exhibits an SB2 orbital period of 4.08 days, $K_1$ = 46.8
km/s, and $K_2$ = 58.1 km/s (elements from Herbig \& Moorhead 1965).
\citet{Davison2014} found the fourth component via radial velocity
monitoring of the C component, determining an SB1 orbit with period of
1.80 days and $K_1$ = 21.4 km/s.  At such a short period, our
astrometry of CD reveals no perturbation, but the proximity of the
components likely causes the high level of photometrically variability
shown in Figure \ref{fig:trend}.  The new D component remains to be
imaged, but the minimum mass is 0.056 $M_{\odot}$, so it is either a
high-mass brown dwarf or M star.

$\bullet$ {\it 2248$-$2422} : {\bf LP 876-010} is the tertiary in the
triple system containing Fomalhaut (A), GJ 879 (B, also known as TW
PsA), and LP 876-010 (C).  LP 876-010 is listed in the middle portion
of Table 1 because A and B were previously known to be within 10 pc,
although the C component was not.  The system was thoroughly discussed
in \citet{Mamajek2013}, where the C component was identified as part
of the system.  The $\pi_{trig}$ in that paper for C (132.07 $\pm$
1.19 mas, updated here to 132.27 $\pm$ 1.07 mas using 10.4 years of
data) from the RECONS program is consistent with the values of 129.81
$\pm$ 0.47 mas for A and 131.42 $\pm$ 0.62 mas for B
\citep{vanLeeuwen2007}.


\section{Discussion}
\label{sec:discussion}

\subsection{The HR Diagram}
\label{sec:hrdiagram}

Figure~\ref{fig:hrdia} illustrates an observational HR diagram for
stars within 25 pc having parallaxes measured during the RECONS
program at the 0.9m, for which a complete list of results can be found
at {\it www.recons.org}.  We use $M_V$ vs. $V-K$ as proxies for the
luminosities and temperatures of targets.

Solid red points represent red and brown dwarfs for which RECONS has
published the first accurate parallax for the system placing it within
10 pc.  Solid blue points represent the two white dwarfs added to the
10 pc sample.  Open circles represent additional stars within 25 pc
for which $\pi_{trig}$ was measured at the 0.9m.  These are primarily
red dwarfs, published in several previous papers in this series
\citep{Jao2005,Henry2006,Riedel2010,Jao2011,Mamajek2013,Jao2014,Dieterich2014,Riedel2014,Lurie2014,Davison2015,Benedict2016,Winters2017,Bartlett2017,Jao2017},
and in two papers reporting parallaxes for white dwarfs
\citep{Subasavage2009,Subasavage2017}.

The new 10 pc systems stretch from L 098-059 and the triple LTT 12352
ABC with $M_V$ = 11.8 and types M2.5V--M3.5V, to the three faintest,
DEN 1048-3956, SCR 1845-6357 AB, and WIS 0720-0846 AB with $M_V$ =
19.4 and types M8.5V--M9.5V.  The single brown dwarf, DEN 0255-4700,
at the lower right of the plot has $M_V$ = 24.4.  The pre-main
sequence star AP Col at 8.6 pc \citep{Riedel2011} is shown with a red
square, clearly elevated above the main sequence red dwarfs.  In all,
27 of the systems have $V-K$ = 5.0--7.0, typical of the red dwarf
population that is rich in spectral types M3.0V--M5.0V.  Among the
nine faintest, reddest, star systems having $V-K$ = 7.0--9.1, there
are five close multiples that are excellent targets for mass
determinations.



\subsection{Evolution of the 10 Parsec Sample}
\label{sec:evolution}

\begin{figure}[ht!]

{\includegraphics[scale=0.70,angle=270]{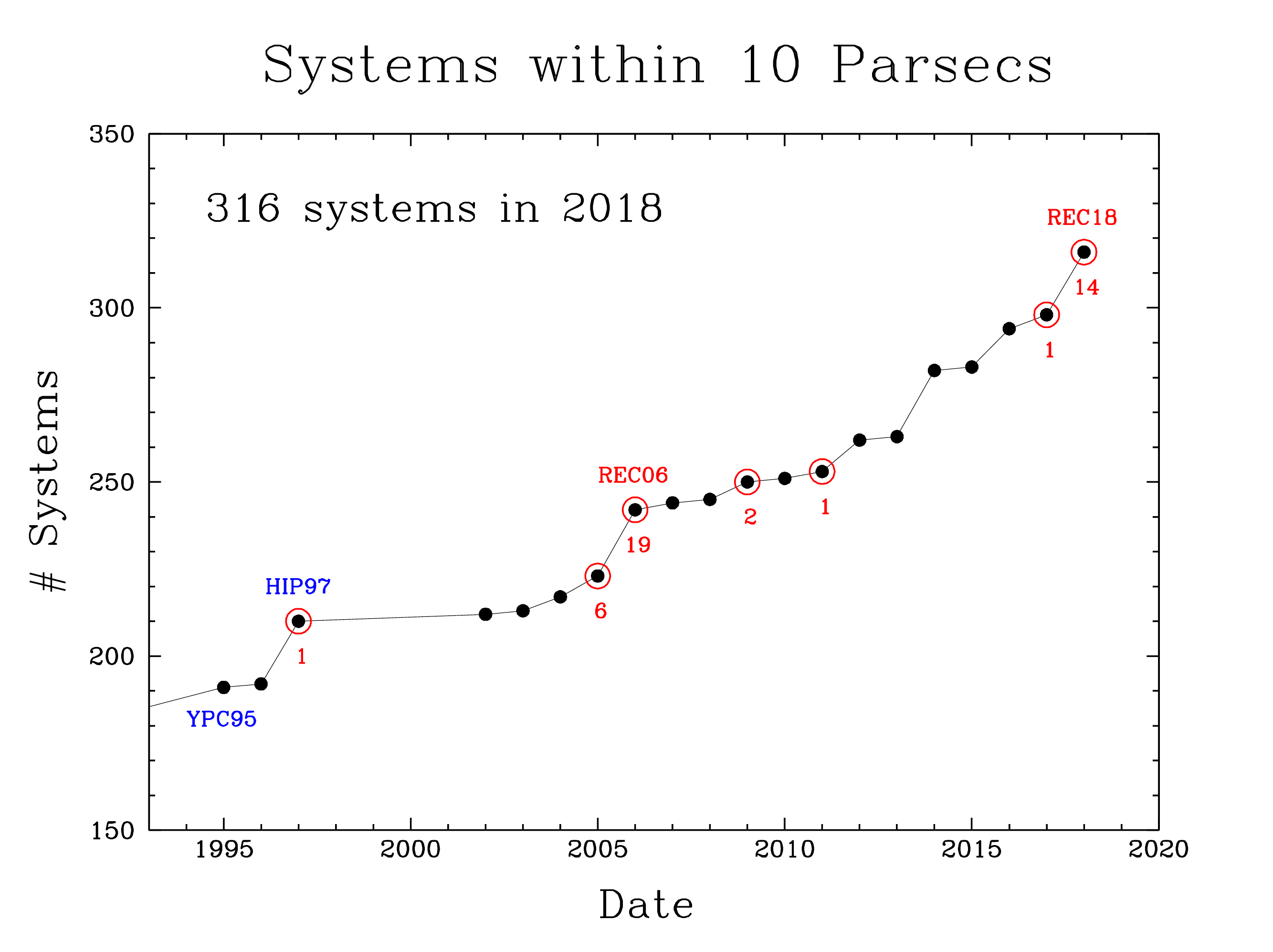}}
\hspace{-0.16in}

\vskip-00pt

\caption{\scriptsize The growth of the 10 pc sample is shown from the
  publication of the Yale Parallax Catalog in 1995 (the YPC95 point)
  to the present.  {\it Hipparcos} contributed 17 systems in 1997, as
  indicated with the label HIP97.  At least one system has been added
  to the sample every year starting in 2002 and the census has
  steadily risen since.  Points encircled with red are years in which
  RECONS added the number of systems indicated with red digits.  REC06
  = \citet{Henry2006} and REC18 = this paper. \label{fig:censu}}

\end{figure}

Figure \ref{fig:censu} charts the total number of systems in the 10 pc
sample over the past two decades, since the publication of the YPC in
1995.  These recent contributions to the 10 pc sample are listed in
Table 4.  Red numbers under various points indicate the 44 systems
added by RECONS to the sample.  After a five-year pause in nearby star
parallaxes after the {\it Hipparcos} mission, the census has continued
to climb steadily, with the additions of stars as well as systems
comprised only of brown dwarfs.  Among the 316 systems known within 10
pc today, only 191 were identified and had parallaxes with errors less
than 10 mas at the time of the YPC, indicating an overall increase of
65\%.  Since 1995, 125 systems have been added, including six FGK
systems from {\it Hipparcos}, three white dwarfs, 79 red dwarf
systems, and 37 brown dwarf systems.  Thus, a total of 88 stellar
systems have been added to the 10 pc sample, more than twice the
number of brown dwarf-only systems.  Notably, the rate of increase has
yet to slow, although of course there will be a turnover in the census
curve when all systems have been found.  We predict that there will be
a short list of stars contributed by {\it Gaia} soon (see
$\S$\ref{sec:gaia}).

Since the millenium, the RECONS group has focused its efforts on
identifying red and white dwarfs missing from the nearby sample
\citep{Henry2003}, and published its first parallaxes in 2005
\citep{Costa2005,Jao2005}, ten years after the YPC.  As outlined in
Table 4, the largest number of systems added to the 10 pc census in
the past two decades was the 18 systems added in
\citet{Henry2006}.\footnote{Two systems near the 10 pc border in that
  paper have now been found to be just further than 10 pc --- APM
  0237-5928 and LHS 1302 --- as discussed in $\S$\ref{sec:notes}.}
The second largest set of additions (17 systems) was from {\it
  Hipparcos} and the third (14 systems) is from this paper.  After
RECONS and {\it Hipparcos}, the largest contributions are the nine
systems from the USNO program in Flagstaff
\citep{Dahn2002,Vrba2004,Dahn2017} and the eight systems from URAT
observations \citep{Finch2016,Finch2018}, for which we provide
confirming parallaxes for seven more systems here.  As detailed in
Table 4, some recent efforts have focused on using infrared arrays to
measures parallaxes for brown dwarfs,\footnote{For many of the brown
  dwarfs, only relative $\pi_{trig}$ have been computed, with no
  corrections to absolute $\pi_{trig}$.  Given the often large errors
  in their parallaxes, the correction is generally much smaller than
  the error, and the overall picture of the solar neighborhood census
  would be incomplete without these new members, so they are included
  here.}  particularly targeting new candidates from the $WISE$
spacecraft.  Nonetheless, somewhat surprising is the continuing
utility of optical CCDs for measuring parallaxes to nearby systems.



\subsection{Missing Systems}
\label{sec:missing}

At this juncture in the RECONS effort to reveal nearby stars, it is
instructive to revisit the estimate of ``missing'' systems outlined in
Figure 1 of \citet{Henry1997}.  For this discussion, we focus on {\it
  stellar} systems, rather than brown dwarfs, given that no brown
dwarf primaries were known in 1997 and they likely remain
underrepresented in the present sample.\footnote{As noted in
  $\S$\ref{sec:evolution}, there are currently 37 systems containing
  only brown dwarfs known within 10 pc.}  In 1997, there were 45
systems known within 5 pc, indicating that there should be 360 systems
within 10 pc, assuming a constant space density.  However, at the
time, only 229 systems were known, implying that 131 systems (36\%)
were unaccounted for.

As of this writing, there have been four {\it stellar}
systems\footnote{Rankings here omit the brown dwarf systems WISE
  1049-5319 AB and UGPS 0722-0540.} found within 5 pc since the
discovery of GJ 1061 (20th nearest), the subject of the
\citet{Henry1997} effort: SO 0253+1652 (23rd nearest), SCR 1845-6357
AB (26th nearest), DEN 1048-3956 (29th nearest), and LHS 288 (43rd
nearest), all of which had their first high-quality parallaxes
published by RECONS.  There has also been a swap in membership since
then, with LP 944-020 dismissed from the 5 pc sample and GJ 682 added
here.  Thus, the gain within 5 pc is four stellar systems, bringing
the total to 49.  Again assuming a constant density, this translates
to 392 systems within 10 pc, while the current 10 pc list contains
only 279 systems with stellar primaries.  This implies that 113
systems (29\%) are still missing.

Given only the slight drop from 36\% incompleteness in 1997 to 29\% in
2018, it seems that we have made only minor progress in completing the
sample, but certainly a 22\% increase (from 229 to 279 systems) in the
number of verified stellar systems nearest to the Sun is a scientific
advance.  However, it is important to note that the standard deviation
of our presumably Poisson sampling of the distribution of the 5 pc
stellar population, $\sigma$ = N$^{1/2}$, where N = 49 star systems,
is $\pm$7 systems.  This corresponds to $\pm$56 systems at 10 pc for
eight times the volume.  Thus, the difference between the observed
(279 systems) and anticipated (392 systems) populations is only twice
the standard deviation, indicating that our progress may be more
significant than perceived if the 392 target systems is an
overestimate.  {\it Considering this numerical treatment and our
  experience in searching for nearby stars over two decades, we
  predict that the complete census of systems within 10 pc with
  stellar primaries will eventually be found to be $\sim$300,
  indicating that the sample is currently at least 90\% complete.}


\subsection{{\it Gaia} Data Releases}
\label{sec:gaia}

On September 14, 2016, the first astrometry data from {\it Gaia} were
provided via Data Release 1 (DR1), including parallaxes for $\sim$2
million stars observed by Tycho, the auxiliary star mapper on the {\it
  Hipparcos} spacecraft.  However, \cite{Jao2016} found small, but
systematic offsets between {\it Gaia} and {\it Hipparcos} parallaxes
in a study of 612 stars within 25 pc.  In addition, a careful
evaluation of the {\it Gaia} DR1 results indicates that there would be
very few changes to the sample.  Of the 62 systems in the 10 pc list
in {\it Gaia} DR1, four --- GJ 239, GJ 678.1, L 026-027, and L 098-059
--- would slip beyond 10 pc, and none by more than a few mas.  In
balance, three new entries would occur --- GJ 172, GJ 436, and TYC
3980-1081-1.  The latter star is the only one for which the parallax
(120.59 $\pm$ 0.96 mas) places it well within 10 pc.  This star was
previously identified as a possible nearby star by \citet{Finch2016}
with $\pi_{trig}$ = 154.8 $\pm$ 12.1 mas, although because of the
large error, it was not yet in the sample.

In sum, {\it Gaia} DR1 data change the character of the 10 pc sample
very little and all of the parallaxes used to determine membership in
the sample are likely to change somewhat in the soon-to-be-released
{\it Gaia} DR2 results.  Thus, instead of incorporating the initial
DR1 results, we await the DR2 results that are based upon a greater
timespan of data, and when the dataset extends to fainter magnitudes.
Ultimately, our goal is to provide a comprehensive view of the 10 pc
sample in a future paper in this series, once the {\it Gaia} DR2
results are available.

%
%

\section{Conclusions}
\label{sec:conclusions}

The RECONS astrometry/photometry program will continue at the 0.9m,
with two primary goals.  First, several hundred of the nearest red
dwarfs are being observed astrometrically to detect unseen companions
in $\sim$decade-long orbits, such as those shown in Figure
\ref{fig:perts}, including a search for Jovian planets.  Second, the
same stars are being observed photometrically to measure their
variability over long timescales to detect cyclic changes in
brightness that mimic our Sun's solar cycle, as shown in Figures
\ref{fig:cycle} and \ref{fig:trend}.

Now is an appropriate time to be circumspect about the accomplishments
of the parallax program to date.  Among the most important are:

$\bullet$ Even after the YPC and {\it Hipparcos} parallax compendia
were available, many of the Sun's neighbors had yet to be revealed.  A
total of 44 systems have been placed within the 10 pc sample via the
RECONS program, including 41 red dwarf systems, two white dwarfs, and
one brown dwarf.  These 44 systems comprise 14\% of all systems
currently known within 10 pc of the Sun.

$\bullet$ Many of the 14 close multiples among the 44 discoveries
promise to yield highly accurate masses.  Because of their proximity,
detailed work on these systems will provide fundamental understanding
of the characteristics of stars (and a few brown dwarfs) that depend
on mass, the all-important parameter that determines nearly everything
that occurs in the life of a star.

$\bullet$ We predict that there are $\sim$300 systems with stellar
primaries within 10 pc, and that the current such sample is at least
90\% complete.

Undoubtedly, some of the Sun's nearest neighbors have yet to be
discovered, but great progress has been made over the past two
decades.  In addition to presenting here the critical parallaxes that
place 44 new systems within 10 pc, we provide additional astrometric,
photometric, and spectroscopic information that allows us to
characterize these systems.  With the list of nearby stars in hand, we
are poised to continue the reconnassiance of the solar neighborhood in
the search for planets orbiting these stars, and ultimately, life on
those planets.

\section{Acknowledgments}

Colleagues at the Cerro Tololo Inter-American Observatory and the
SMARTS Consortium have played integral parts in supporting the RECONS
effort over two decades at CTIO and we are indebted to all those at
CTIO and SMARTS who have made this work possible.  At CTIO, Alberto
Miranda and Joselino Vasquez made a large number of observations at
the 0.9m over many years.  Mountain operations have been supported by
arguably the best observatory staff in the world, and we wish to thank
Marco Bonati, Edgardo Cosgrove, Arturo Gomez, Manuel Hernandez,
Rodrigo Hernandez, Peter Moore, Humberto Orrego, Esteban Parkes, David
Rojas, Javier Rojas, Mauricio Rojas, Oscar Saa, Hernan Tirado, Ricardo
Venegas, and so many others who have provided superb and enduring
assistance with the telescope, instrument, and computer systems.  Many
students and colleagues have also been a part of this long-term work,
and we are particularly indebted to Mark Boyd, Misty Brown, Edgardo
Costa, Tiffany Clements, Cassy Davison (Smith), Altonio Hosey, John
Lurie, Rene Mendez, Hektor Monteiro, Leonardo Paredes, and R.~Andrew
Sevrinsky.  We also thank Norbert Zacharias, who was instrumental in
the URAT effort for which seven supporting parallaxes are included
here.

The earliest phase of the RECONS astrometry/photometry program was
supported by the NASA/NSF Nearby Star (NStars) Project through NASA
Ames Research Center.  The National Science Foundation has been
consistently supportive of this effort under grants AST-0507711,
AST-0908402, AST-1109445, AST-141206, and AST-1715551, while for
several years the effort was also supported by NASA's {\it Space
  Interferometry Mission}.

This work has used data products from the Two Micron All Sky Survey,
which is a joint project of the University of Massachusetts and the
Infrared Processing and Analysis Center at California Institute of
Technology funded by NASA and NSF.  Information was collected from
several additional large database efforts: the SIMBAD database and the
VizieR catalogue access tool, operated at CDS, Strasbourg, France,
NASA's Astrophysics Data System, the SuperCOSMOS Science Archive,
prepared and hosted by the Wide Field Astronomy Unit, Institute for
Astronomy, University of Edinburgh, which is funded by the UK Science
and Technology Facilities Council, and the Washington Double Star
Catalog maintained at the U.S. Naval Observatory.

%


\clearpage

\begin{deluxetable}{llccccccccccrrrrrcl}
\voffset000pt
\rotate
\setlength{\tabcolsep}{0.03in}
\label{tab:astr}
\tablewidth{0pt}
\tabletypesize{\tiny}
\tablecaption{Astrometric Results}
\tablehead{\colhead{Name}                &
           \colhead{Other}               &
           \colhead{R.A.}                &
           \colhead{Decl.}               &
           \colhead{Disc.}               &
           \colhead{Time}                &
           \colhead{Filter}              &
           \colhead{N$_{sea}$}           &
           \colhead{N$_{frm}$}           &
           \colhead{Coverage}            &
           \colhead{Time}                &
           \colhead{N$_{ref}$}           &
           \colhead{$\pi$(rel)}          &
           \colhead{$\pi$(corr)}         &
           \colhead{$\pi$(abs)}          &
           \colhead{$\mu$}               &
           \colhead{$\theta$}            &
           \colhead{$V_{tan}$}           &
           \colhead{notes}               \\

           \colhead{}                    &
           \colhead{Name}                &
           \colhead{J2000.0}             &
           \colhead{J2000.0}             &
           \colhead{Ref.}                &
           \colhead{yr}                  &
           \colhead{}                    &
           \colhead{}                    &
           \colhead{}                    &
           \colhead{}                    &
           \colhead{yr}                  &
           \colhead{}                    &
           \colhead{mas}                 &
           \colhead{mas}                 &
           \colhead{mas}                 &
           \colhead{mas yr$^{-1}$}       &
           \colhead{deg}                 &
           \colhead{km sec$^{-1}$}       &
           \colhead{}                    \\

           \colhead{(1)}                 &
           \colhead{(2)}                 &
           \colhead{(3)}                 &
           \colhead{(4)}                 &
           \colhead{(5)}                 &
           \colhead{(6)}                 &
           \colhead{(7)}                 &
           \colhead{(8)}                 &
           \colhead{(9)}                 &
           \colhead{(10)}                &
           \colhead{(11)}                &
           \colhead{(12)}                &
           \colhead{(13)}                &
           \colhead{(14)}                &
           \colhead{(15)}                &
           \colhead{(16)}                &
           \colhead{(17)}                &
           \colhead{(18)}                &
           \colhead{(19)}                }
\startdata
\multicolumn{19}{c}{New 10 Parsec Members --- 44 systems} \\ 
\hline
\\
GJ 2005 ABC       & LHS 1070 ABC  & 00 24 44.19 & $-$27 08 24.2 & Cos05 &  2.30 & R & 19s & 172 & 1999.64$-$2017.93 & 18.30 &  6 &  138.32$\pm$1.59 &  0.98$\pm$0.04 &  139.30$\pm$1.59 &  687.3$\pm$0.3 &  350.5$\pm$0.04 &  23.4 & orbit fit                         \\  
WD 0038-226       & GJ 2012       & 00 41 26.03 & $-$22 21 02.3 & Sub09 &  8.25 & R & 18s & 168 & 1999.64$-$2016.97 & 17.33 &  7 &  109.60$\pm$0.76 &  1.24$\pm$0.07 &  110.84$\pm$0.76 &  608.2$\pm$0.1 &  232.6$\pm$0.02 &  26.0 &                                   \\  
WD 0141-675       & LHS 145       & 01 43 00.98 & $-$67 18 30.4 & Sub09 &  7.42 & V & 12s & 163 & 2000.57$-$2015.82 & 15.25 &  6 &  100.92$\pm$0.79 &  0.88$\pm$0.07 &  101.80$\pm$0.79 & 1079.6$\pm$0.2 &  199.2$\pm$0.01 &  50.3 & Sub17 result                      \\  
L 173-019         &               & 02 00 38.31 & $-$55 58 04.8 &  new  &\nodata& V &  9s & 120 & 2007.81$-$2015.97 &  8.16 &  7 &  122.77$\pm$2.07 &  0.34$\pm$0.18 &  123.11$\pm$2.08 &  125.5$\pm$0.8 &  123.5$\pm$0.74 &   4.8 &                                   \\  
SO 0253+1652      &               & 02 53 00.89 & $+$16 52 52.7 & Hen06 &  2.35 & I & 13s & 165 & 2003.53$-$2015.83 & 12.30 &  6 &  258.25$\pm$1.73 &  2.76$\pm$0.32 &  261.01$\pm$1.76 & 5103.1$\pm$0.5 &  138.1$\pm$0.01 &  92.7 &                                   \\  
DENIS 0255-4700   &               & 02 55 03.68 & $-$47 00 51.6 & Cos06 &  3.20 & I & ... &  44 &        ...        &  3.20 & 17 &  201.14$\pm$3.89 &  0.23$\pm$0.05 &  201.37$\pm$3.89 & 1148.5$\pm$2.2 &  119.5$\pm$0.21 &  27.0 & Cos06 result (1.5m)               \\
\enddata

\tablecomments{References:
And11 = \citet{Andrei2011},
Bar17 = \citet{Bartlett2017},
Cos05 = \citet{Costa2005},
Cos06 = \citet{Costa2006},
Die14 = \citet{Dieterich2014},
Dit14 = \citet{Dittmann2014},
Dup12 = \citet{Dupuy2012},
Fah12 = \citet{Faherty2012},
Fin16 = \citet{Finch2016},
HIP97 = \citet{ESA1997},
Hen06 = \citet{Henry2006},
Hen97 = \citet{Henry1997},
Jao05 = \citet{Jao2005},
Rie11 = \citet{Riedel2011},
Sah14 = \citet{Sahlmann2014},
Sub09 = \citet{Subasavage2009},
Sub17 = \citet{Subasavage2017},
Tin96 = \citet{Tinney1996},
Wei16 = \citet{Weinberger2016},
YPC95 = \citet{vanAltena1995},
}
\end{deluxetable}

\voffset000pt{
\hoffset000pt{
\begin{deluxetable}{lcrrrrrrrr}
\setlength{\tabcolsep}{0.03in}
\label{tab:photspec}
\tablewidth{0pt}
\tabletypesize{\tiny}
\tablecaption{Photocentric Orbits for Nearby Systems}
\tablehead{\colhead{Name}                &
           \colhead{\#orbits}            &
           \colhead{$P$}                 &
           \colhead{$P$}                 &
           \colhead{$a$}                 &
           \colhead{$e$}                 &
           \colhead{$i$}                 &
           \colhead{$\omega$}            &
           \colhead{$\Omega$}            &
           \colhead{$T_0$}               \\

           \colhead{}                    &
           \colhead{}                    &
           \colhead{(yr)}                &
           \colhead{(d)}                 &
           \colhead{(mas)}               &
           \colhead{}                    &
           \colhead{(deg)}               &
           \colhead{(deg)}               &
           \colhead{(deg)}               &
           \colhead{}                    \\

           \colhead{(1)}                 &
           \colhead{(2)}                 &
           \colhead{(3)}                 &
           \colhead{(4)}                 &
           \colhead{(5)}                 &
           \colhead{(6)}                 &
           \colhead{(7)}                 &
           \colhead{(8)}                 &
           \colhead{(9)}                 }
\startdata
                  &        &          &          &        &        &         &         &         &           \\ %
2MA 1750-0016 AB  &   1.2  &    6.74  &  2460.4  &  12.7  &  0.08  &  110.0  &  339.0  &   21.5  &  2024.04  \\ %
                  &        &    0.39  &   143.5  &   1.2  &  0.09  &    7.3  &   83.4  &    8.3  &     1.60  \\ %
                  &        &          &          &        &        &         &         &         &           \\ %
GJ 54 AB          &  15.1  &    1.15  &   419.1  &  18.7  &  0.04  &  124.2  &  102.0  &  102.3  &  2018.75  \\ %
                  &        & $<$0.01  &     1.3  &   1.9  &  0.05  &    7.7  &   73.7  &    9.7  &     0.24  \\ %
                  &        &          &          &        &        &         &         &         &           \\ %
GJ 595 AB         &  31.8  &    0.17  &    62.0  &   9.0  &  0.02  &   94.5  &  217.9  &  205.9  &  2018.43  \\
\enddata



\end{deluxetable}
}
}




%


\begin{deluxetable}{lccrcrcrcccrrcrcrclccl}
\hoffset-30pt
\setlength{\tabcolsep}{0.03in}
\label{tab:photspec}
\tablewidth{0pt}
\tabletypesize{\tiny}
\tablecaption{Photometric and Spectroscopic Results}
\tablehead{\colhead{Name}                &
           \colhead{R.A.}                &
           \colhead{Decl.}               &
           \colhead{$V$}                 &
           \colhead{ }                   &
           \colhead{$R$}                 &
           \colhead{ }                   &
           \colhead{$I$}                 &
           \colhead{ }                   &
           \colhead{\#}                  &
           \colhead{$\pi$}               &
           \colhead{$\sigma$}            &
           \colhead{$J$}                 &
           \colhead{ }                   &
           \colhead{$H$}                 &
           \colhead{ }                   &
           \colhead{$K$}                 &
           \colhead{ }                   &
           \colhead{Spec}                &
           \colhead{Lum}                 &
           \colhead{Spec}                &
           \colhead{notes}               \\

           \colhead{}                    &
           \colhead{J2000.0}             &
           \colhead{J2000.0}             &
           \colhead{mag}                 &
           \colhead{ }                   &
           \colhead{mag}                 &
           \colhead{ }                   &
           \colhead{mag}                 &
           \colhead{ }                   &
           \colhead{nts}                 &
           \colhead{filter}              &
           \colhead{mmag}                &
           \colhead{mag}                 &
           \colhead{ }                   &
           \colhead{mag}                 &
           \colhead{ }                   &
           \colhead{mag}                 &
           \colhead{ }                   &
           \colhead{Type}                &
           \colhead{Class}               &
           \colhead{ref}                 &
           \colhead{}                    \\

           \colhead{(1)}                 &
           \colhead{(2)}                 &
           \colhead{(3)}                 &
           \colhead{(4)}                 &
           \colhead{ }                   &
           \colhead{(5)}                 &
           \colhead{ }                   &
           \colhead{(6)}                 &
           \colhead{ }                   &
           \colhead{(7)}                 &
           \colhead{(8)}                 &
           \colhead{(9)}                 &
           \colhead{(10)}                &
           \colhead{ }                   &
           \colhead{(11)}                &
           \colhead{ }                   &
           \colhead{(12)}                &
           \colhead{ }                   &
           \colhead{(13)}                &
           \colhead{(14)}                &
           \colhead{(15)}                &
           \colhead{(16)}                }
\startdata
\multicolumn{22}{c}{New 10 Parsec Members --- 44 systems} \\ 
\hline
                  &              &               &         &   &       &   &       &   &     &   &      &        &   &       &   &       &   &      &     &       &                        \\ 
GJ 2005 ABC       &  00 24 44.19 & $-$27 08 24.2 &   15.28 & J & 13.52 & J & 11.37 & J & 665 & R & 17.2 &   9.25 & J &  8.55 & J &  8.24 & J & M5.5 & VJ  &   *   & ABC $<$2\arcsec        \\ 
WD 0038-226       &  00 41 26.03 & $-$22 21 02.3 &   14.50 &   & 14.08 &   & 13.67 &   & 444 & R &  7.4 &  13.34 &   & 13.48 &   & 13.74 &   & DQ   & pec & Sub17 &                        \\ 
WD 0141-675       &  01 43 00.98 & $-$67 18 30.4 &   13.82 &   & 13.52 &   & 13.23 &   & 333 & V &  6.4 &  12.87 &   & 12.66 &   & 12.58 &   & DAZ  &     & Sub17 &                        \\ 
L 173-019         &  02 00 38.31 & $-$55 58 04.8 &   11.90 &   & 10.70 &   &  9.15 &   & 333 & V & 19.9 &   7.63 &   &  7.09 &   &  6.77 &   & M2.0 & V   &   *   &                        \\ 
SO 0253+1652      &  02 53 00.89 & $+$16 52 52.7 &   15.14 &   & 13.03 &   & 10.65 &   & 333 & I & 10.3 &   8.39 &   &  7.88 &   &  7.59 &   & M6.5 & V   &   *   &                        \\ 
DEN 0255-4700     &  02 55 03.68 & $-$47 00 51.6 &   22.92 &   & 19.91 &   & 17.45 &   & 111 & I &\nodata& 13.25 &   & 12.20 &   & 11.56 &   & L8   &     & Fil15 & from Cos06             \\
\enddata

\tablecomments{References: 
* = this paper,
Bur15 = \citet{Burgasser2015}, 
Cos06 = \citet{Costa2006}, 
Fil15 = \citet{Filippazzo2015},
Giz02 = \citet{Gizis2002}, 
Ham04 = \citet{Hambaryan2004}, 
Ken07 = \citet{Kendall2007}, 
Per14 = \citet{Perez2014}, 
Raj13 = \citet{Rajpurohit2013}, 
Rei08 = \citet{Reid2008}, 
Sch10 = \citet{Schmidt2010}
Sub17 = \citet{Subasavage2017}
}

\tablecomments{A ``J'' next to a magnitude or spectral type indicates
  that more than one object is included in the measurement value,
  i.e., a close multiple system. }


\end{deluxetable}





\centering
\begin{deluxetable}{lcccccl}
\voffset50pt
\rotate
\setlength{\tabcolsep}{0.03in}
\label{tab:sysadd}
\tablewidth{0pt}
\tabletypesize{\tiny}
\tablecaption{Systems Added to the 10 Parsec Census Since 1995.\tablenotemark{a}}
\tablehead{
	   \colhead{Reference}                       &
	   \colhead{Observing}                       &
	   \colhead{Other\tablenotemark{b}}          &
	   \colhead{Red\tablenotemark{b}}            &
	   \colhead{Brown\tablenotemark{b}}          &
 	   \colhead{Total}                           &
 	   \colhead{New 10 pc}                       \\

           \colhead{}                                &
           \colhead{Method}                          &
           \colhead{Dwarfs}                          &
           \colhead{Dwarfs}                          &
           \colhead{Dwarfs}                          &
           \colhead{Systems}                         &
           \colhead{Systems}                         }
\startdata
\multicolumn{7}{c}{New Systems within 10 Parsecs from RECONS Parallax Efforts}                                                                         \\
\hline                                                                                                                                                   
                                                                                                                                                       \\
Henry et al.~(1997)         & phot plates          &  0 &  1 &  0 &  1 & GJ1061                                                                        \\
Costa et al.~(2005)         & optical CCD          &  0 &  1 &  0 &  1 & GJ2005                                                                        \\ 
Jao et al.~(2005)           & optical CCD          &  0 &  5 &  0 &  5 & GJ754 GJ1068 GJ1123 GJ1128 DEN1048-3956                                       \\
Costa et al.~(2006)         & optical CCD          &  0 &  0 &  1 &  1 & DEN0255-4700                                                                  \\
Henry et al.~(2006)         & optical CCD          &  0 & 18 &  0 & 18 & see Table 1                                                                   \\ 
Subasavage et al.~(2009)    & optical CCD          &  2 &  0 &  0 &  2 & WD0038-226 WD0141-675 (2 white dwarf systems)                                 \\
\enddata

\tablenotetext{a}{Systems must have weighted mean trigonometric parallaxes of 100 mas or more and errors of 10 mas or less.}
\tablenotetext{b}{Primaries are used to categorize systems as white, red, or brown dwarf additions.}
\end{deluxetable}

%

\clearpage


\clearpage


\clearpage


\clearpage


\clearpage


\clearpage


\clearpage


\end{document}